\documentclass[12pt]{article}
\usepackage{amsmath,amssymb,amsthm,booktabs}
\usepackage{mathrsfs}
\usepackage{geometry}
\usepackage{pifont}
\usepackage[OT1]{fontenc}
\usepackage[utf8]{inputenc}
\usepackage[colorlinks,citecolor=blue,urlcolor=blue]{hyperref}
\usepackage{txfonts}
\usepackage{indentfirst}
\usepackage{multirow}
\usepackage{float,subfig}

\usepackage{bm}
\usepackage{euscript}
\usepackage{graphicx}
\usepackage{multicol}
\usepackage[usenames,dvipsnames,svgnames,table]{xcolor}

\usepackage[round]{natbib}
\usepackage[ruled]{algorithm2e}

\numberwithin{equation}{section} \theoremstyle{plain}
\newtheorem{theorem}{Theorem}[section]
\newtheorem{lemma}{Lemma}[section]
\newtheorem{corollary}{Corollary}[section]

\newtheorem{example}{Example}[section]
\newtheorem{definition}{Definition}[section]
\newtheorem{remark}{Remark}[section]

\begin{document}

\newcommand{\gai}[1]{{#1}}


\makeatletter
\def\ps@pprintTitle{%
  \let\@oddhead\@empty
  \let\@evenhead\@empty
  \let\@oddfoot\@empty
  \let\@evenfoot\@oddfoot
}
\makeatother

\newcommand\tabfig[1]{\vskip5mm \centerline{\textsc{Insert #1 around here}}  \vskip5mm}

\vskip2cm

\title{Asset price bubbles under model uncertainty and short-sale constraints: A discrete-time analysis}
\author {Wenqing Zhang\thanks{Corresponding author, College of Mathematics and Physics, China Three Gorges University, PR China, (wqzhang1025@163.com).}
\quad Lin Zhang\thanks{Center for Inclusive Finance and Agricultural \& Rural Development Research \& School of Economics and Management, Southwest University, PR China, (zhanglin2762@126.com).}
}
\date{}
\maketitle

\begin{abstract}
In this study, we investigate asset price bubbles in a discrete-time, discrete-state market under model uncertainty and short-sale constraints. Using a super-hedging valuation benchmark, we study the difference between the asset's market price and its fundamental value in this constrained market. We examine how assumptions on the liquidation time affect the conditional expectation bounds satisfied by the resulting bubble process. For assets with bounded maturity and no dividend payments, the $G$-supermartingale property of prices provides a necessary and sufficient condition for the existence of bubbles. In contrast, when maturity is unbounded, the infi-supermartingale property yields a necessary condition, while the $G$-supermartingale property remains sufficient. We also show that no dominance rules out bubbles when the liquidation time is bounded. For finite-horizon contingent claims, we distinguish upper-valuation relations from no-dominance benchmarks. Fundamental prices satisfy put-call bounds, whereas market prices satisfy put-call parity under no dominance. In the dividend-free setting, the same assumptions imply equality between American and European call fundamental values and, under the stated trading conditions, between their market prices.
\end{abstract}


\section{Introduction}
\label{sec:introduce}

Asset price bubbles raise a basic valuation question: when does an asset's market price exceed the value justified by its future cash flows? Answering this question requires both a valuation of these cash flows and an account of the trading restrictions faced by investors. 
Under model uncertainty, the cash-flow valuation depends on the choice of probability model. At the same time, short-sale constraints may prevent investors from exploiting an apparent overvaluation through an admissible trading strategy. It is therefore necessary to examine how fundamental values and price bubbles should be defined when model uncertainty and short-sale constraints are considered jointly.

Classical episodes, such as the Dutch Tulipmania, the Mississippi Bubble, and the South Sea Bubble, have been extensively documented and analyzed \citep{Garber1990famous}.
More recently, the bursting of the housing price bubble and its connection to the subprime mortgage crisis has renewed interest in bubble phenomena within both the financial industry and academic research \citep{Jarrow2009forward,Jarrow2011detect}.
In general terms, a bubble is a deviation between the trading price of an asset and its underlying value.
The economic literature has examined this deviation from various perspectives in an attempt to explain why and under what conditions bubbles arise.
For example, \citet{Harrison1978speculative} showed that speculative trading may cause prices to persistently exceed fundamental values.
Other mechanisms contributing to bubble formation include traders' myopic behavior \citep{Tirole1982possibility}, overconfidence \citep{Scheinkman2003overconfidence}, trend-chasing behavior \citep{Follmer2005equilibria}, and the presence of noise traders \citep{Delong1990noise}.

Recent research has further developed the economic theory of asset price bubbles. \citet{HiranoToda2024BubbleEconomics} provide an overview of rational bubbles, with particular attention to dividend-paying assets such as stocks, housing, and land.
\citet{HiranoToda2025BubbleNecessity} establish conditions under which all equilibria are bubbly, relating the necessity of bubbles to economic growth, dividend growth, and interest rates.
These contributions examine the economic conditions supporting bubbles in equilibrium and provide a broader context for the valuation-based analysis pursued here.

In recent decades, a growing literature has leveraged tools from mathematical finance, particularly the theory of (local) martingales, to analyze asset price bubbles.
The martingale approach originated from the seminal work of \citet{Loewenstein2000rational}, who introduced a probabilistic framework for studying bubbles, and was further advanced by \citet{Cox2005local}.
Subsequent contributions include \citet{Jarrow2010}, who considered markets with time-varying local martingale measures, leading to results that differ markedly from classical theory, and \citet{Biagini2014shifting}, who analyzed a flow of equivalent martingale measures under which an asset price initially behaves as a submartingale, later becomes a supermartingale, and eventually returns to zero.
Parallel to these developments, numerous studies investigate bubbles across various asset classes, including forwards and futures \citep{Jarrow2009forward}, equities \citep{Jarrow2011there}, foreign exchange \citep{Jarrow2011foreign}, bonds \citep{Bilina2016relative}, options \citep{Jarrow2021inferring}, and so on \citep{Jarrow2016testing}.

Recent studies have also examined how option prices can be used to identify asset price bubbles.
\citet{FusariJarrowLamichhane2025OptionsData} develop a method that exploits the differential pricing of put and call options.
\citet{StahlBlauth2024MartingaleDefects} investigate martingale defects inferred from option volatility surfaces and their association with bubble conditions in the underlying stock market.
\citet{BiaginiEtAl2025DeepLearning} develop a deep learning approach to bubble detection using observed call option prices.
These studies demonstrate the relevance of derivative prices for the analysis of bubbles. They also motivate examining how the relationship between underlying assets and contingent claims depends on the valuation framework and the trading restrictions.

Discrete-time frameworks also play a prominent role in the study of financial markets, owing to their structural simplicity and their closer correspondence to empirically observed price dynamics.
Their tractability makes them especially suitable for modeling periodic collapses and drawdowns characteristic of bubble behavior \citep{Schatz2020inefficient}. 
From the perspective of financial mathematics, discrete-time models require special attention because strict local martingales and singular processes, central to many continuous-time bubble models, cannot arise; in discrete time, any nonnegative local martingale is necessarily a true martingale \citep{Jarrow2012discrete,Herdegen2022bubbles}.
This necessitates a separate investigation of bubble phenomena in discrete time.
Indeed, \citet{Santos1997rational} provided a systematic analysis of the conditions under which rational bubbles emerge in competitive equilibria; 
\citet{Fukuta1998simple} studied incomplete bursting bubbles and their interrelationships;
and \citet{Herdegen2022bubbles} proposed a new definition of bubbles in discrete-time models based on loss of mass in discounted asset prices.

In addition to model structure, short sales restrictions play a significant role in financial markets, as they help stabilize prices during financial crises and are widely implemented in most emerging economies \citep{Pulido14}. 
However, short-sale constraints may also facilitate the formation and persistence of asset price bubbles by impeding arbitrage and contributing to overvaluation \citep{Miller1977risk,Jarrow2019capital}, and the persistence of such bubbles has been explored in \citet{Lim2011short}.
In addition, \citet{Hong2006asset} examined the relationship between asset float and bubbles under short-sale constraints; 
and \citet{Kocherlakota2008injecting} constructed equilibrium allocations exhibiting bubbles induced by short sales limitations.
Moreover, in many continuous-time models, admissibility conditions requiring wealth processes to remain bounded from below effectively impose implicit short-sale constraints, permitting bubble phenomena in otherwise arbitrage-free settings \citep{Jarrow2010}.

Despite the extensive literature on bubbles, relatively little attention has been paid to Knightian uncertainty, an intrinsic feature of modern financial markets.
Knightian uncertainty can invalidate the classical risk-neutral pricing framework, potentially causing abrupt asset price movements without corresponding changes in fundamentals \citep{Epstein1995uncertainty}.
To model such uncertainty, \citet{Peng1997,Peng2004,Peng2006,Peng2008,Peng2019} introduced sublinear expectation theory, replacing the traditional single probability measure with a family of probability measures, thereby capturing both mean and volatility uncertainty.
This framework has since been widely applied in financial modeling \citep{EJ13,EJ14,Peng2022,Peng2023}.
Within the literature on asset price bubbles, \citet{Biagini2017} study robust fundamental values under model uncertainty and their relation to superreplication.

Trading restrictions also have an established role in robust valuation. \citet{CoxHouObloj2016} study robust pricing and hedging under short-selling restrictions, including supermartingale valuation measures and connections with financial bubbles. Our analysis focuses on an asset with intermediate cash flows and a possibly random liquidation time in a discrete-time, discrete-state market. The questions concern the resulting valuation difference, its dynamics under different liquidation-time assumptions, and the restrictions imposed by no dominance.

Building on this literature, we examine the valuation of future cash flows and the characterization of bubbles in a discrete-time market with model uncertainty and explicit short-sale constraints. 
The joint treatment of these features is relevant because the valuation benchmark must account for the family of probability models, while the assessment of arbitrage must respect the restrictions on trading. This raises three related questions. 
How can a fundamental value based on constrained super-hedging be used to define an asset price bubble under model uncertainty? 
How does the asset's maturity affect the properties of the resulting bubble process? 
What implications does this framework have for the valuation of contingent claims and the price relationships between different contracts?
These questions connect the analysis of bubbles in the underlying asset with the study of forward contracts and European and American options.

We consider a discrete sample space $\Omega=\{\omega_1,\cdots, \omega_K\}$ with $K\in\mathbb{Z}_{+}$ states, where $\mathbb{Z}_{+}$ denotes the set of positive integers. Model uncertainty is represented by a family of probability measures $\mathcal{P}$ defined on $\Omega$. We further assume that $\sup_{P\in\mathcal{P}} P(\omega)>0$ for all $\omega\in\Omega$. 
Time evolves on the discrete horizon $[0,\bar{T}]$, where $\bar{T}$ is either a finite integer $N$ or $\infty$.
For the traded asset, we introduce the discounted wealth process $W$ associated with the discounted market price $\hat{S}_t$, the discounted dividend process $\hat{D}_t$ and the discounted terminal payoff $\hat{X}_{\tau}$. Specifically,
$$
W_t = \hat{S}_t I_{\{t<\tau\}} + \sum_{u=0}^{t\wedge\tau} \hat{D}_u + \hat{X}_{\tau} I_{\{\tau\le t\}}.
$$
Let $\pi$ denote the risky-asset position, which is restricted to be nonnegative due to the short selling prohibition.
The money-market position is unrestricted, and the discounted self-financing wealth satisfies
$V_t=V_0+\sum_{k=1}^{t}\pi_k(W_k-W_{k-1})$.
Within this framework, we establish a no-arbitrage pricing under model uncertainty and short-sale constraints. We show that the no-arbitrage condition holds requires that the existence of a strong risk neutral nonlinear expectation, that is 
$$
W_t\ge \sup_{Q\in\mathcal{Q}} E_Q[W_u \mid \mathcal{F}_t],\quad 0\le t\le u,\quad \sup_{Q\in \mathcal{Q}}Q(\omega)>0,\ \forall \omega\in\Omega.
$$
Building on this result, we derive a super-hedging theorem under model uncertainty and short-sale constraints, demonstrating that the minimal super-hedging price of a contingent claim is given by the robust supremum expectation $\sup_{Q\in\mathcal{Q}} E_Q[H]$.

\begin{figure}[htbp]
    \centering
    \includegraphics[width=\linewidth]{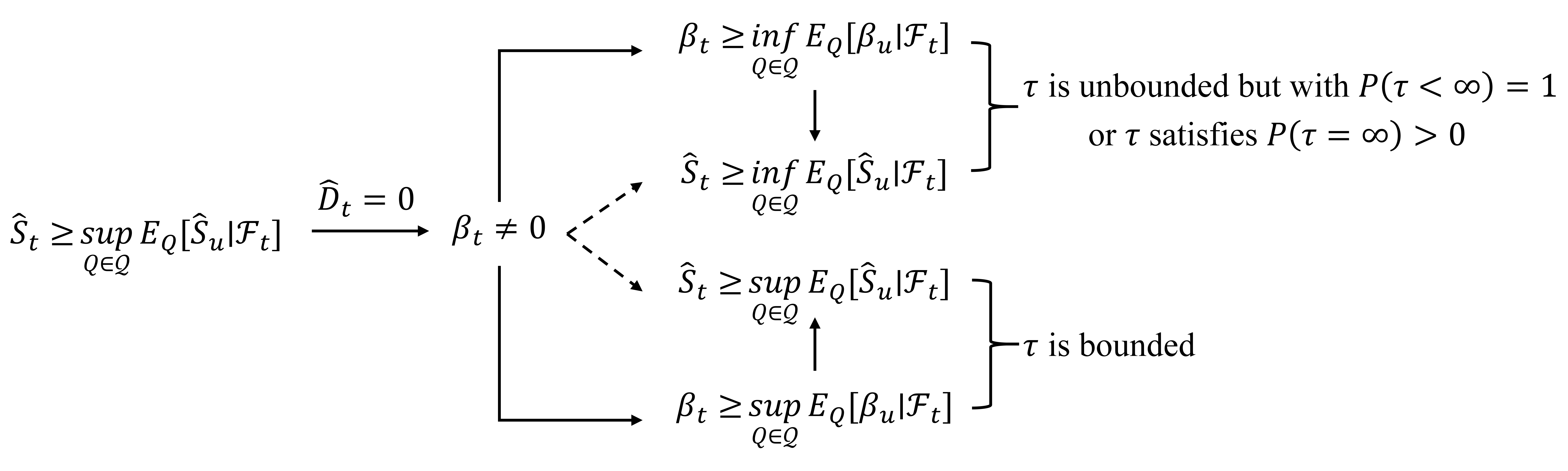}
    \caption{The necessary and sufficient conditions for bubbles}
    \label{fig:Bubble}
\end{figure}

Following the super-hedging approach to fundamental valuation, we use $S^*$ to value the asset's remaining cash flows under the admissible trading restrictions. The corresponding upper-expectation representation is
$$
S_t^* = \sup_{Q\in\mathcal{Q}} E_Q \left[\sum_{u=t+1}^{\tau} \hat{D}_u + \hat{X}_{\tau} I_{\{\tau<\infty\}} \ \middle| \ \mathcal{F}_t \right].
$$
We first establish that this definition is well-defined, and then analyze its asymptotic behavior.
In particular, we show that the process $(S_t^*)_{t\ge 0}$ converges to $0$ q.s.
Using this result, we derive the convergence properties of the associated fundamental wealth process $W^*$ and identify its $G$-martingale structure.
The asset price bubble $\beta$ is then defined as the difference between the market price and the fundamental price:
$
\beta_t = \hat{S}_t - S_t^*.
$
We study the conditional expectation bounds on the bubble process under different assumptions on the liquidation time $\tau$. If $\tau$ is bounded, then $\beta_t$ is a $G$-supermartingale, that is,
$
\beta_t \ge \sup_{Q\in\mathcal{Q}} E_Q[\beta_u \mid \mathcal{F}_t],\quad t\le u.
$
When $\tau$ is unbounded, either allowing $P(\tau=\infty)>0$ or satisfying $P(\tau<\infty)=1$ while remaining unbounded, the bubble process $\beta_t$ instead satisfies an infi-supermartingale property, that is, $
\beta_t\ge \inf_{Q\in\mathcal{Q}} E_Q[\beta_u \mid \mathcal{F}_t],\quad t\le u.
$
Several illustrative examples are provided to clarify these behaviors.
We then investigate structural conditions on market prices that characterize the existence of bubbles.
When $\tau$ is bounded and the asset pays no dividends, we show that the $G$-supermartingale property of the asset is both necessary and sufficient for a bubble to arise.
In contrast, when $\tau$ is unbounded, the infi-supermartingale condition is necessary, while the $G$-supermartingale condition becomes sufficient. 
These results are illustrated in Figure \ref{fig:Bubble}.
They lead to the following key properties of bubbles: (i) $\beta_t\ge 0$ for all $t$; (ii) $\beta_{\tau} I_{\{\tau<\infty\}} = 0$; (iii) under bounded $\tau$, if $\beta_t=0$ then $\beta_u=0$ for all $u\ge t$, while this persistence need not hold when $\tau$ is unbounded. We further show that, when the liquidation time is bounded, no dominance rules out asset price bubbles.

We then study contingent claims over a fixed finite horizon $[0,T]$, assuming that the underlying asset pays no dividends and satisfies $\tau>T$ quasi-surely. In this section, the underlying fundamental value is defined using its resale value at $T$, while retaining the notation $S_t^*$ and $\beta_t$ and suppressing their dependence on the fixed horizon. We first derive bounds for the fundamental prices of European calls and puts; these bounds follow from upper valuation and do not require no dominance. When no dominance is imposed in the enlarged market, market put-call parity holds, the underlying and forward valuation differences vanish, and American and European call fundamental values coincide. Under the stated dynamic trading assumptions, their market prices also coincide. These results provide a no-dominance benchmark rather than a mechanism generating a positive bubble-related early-exercise premium.

The main contributions of this paper are threefold:

(i). We formulate a no-arbitrage criterion and a super-hedging representation for the discrete-time market with model uncertainty and short-sale constraints. 
In a discrete time and states framework, we prove that the $G$-supermartingale property of the wealth process $W$ is the sufficient condition the no-arbitrage condition under model uncertainty and short-sale constraints.
Moreover, the minimal super-hedging cost of a contingent claim is given by the supremum expectation over the set of probability measures $\mathcal{Q}$.

(ii). Using the super-hedging valuation benchmark, we study the fundamental value of the asset's remaining cash flows and the associated bubble process. We analyze their convergence properties and the conditional expectation bounds associated with different assumptions on the liquidation time. We also identify the conditions under which no dominance forces equality between the market price and the super-hedging value of the payoff used in the portfolio comparison.

(iii). We distinguish general upper-valuation relations for contingent claims from the additional restrictions imposed by no dominance. We derive put-call bounds for fundamental prices and put-call parity for market prices. In the dividend-free finite-horizon setting, the super-hedging representation and no dominance yield equality between American and European call fundamental values, and between their market prices under the stated trading assumptions.

The remainder of this paper is organized as follows.
Section \ref{sec:basic} introduces the basic financial market framework and establishes the no-arbitrage pricing together with the super-hedging theorem under model uncertainty and short-sale constraints.
After providing the fundamental price of asset in the concept of asset bubbles, we provides a detailed analysis of bubbles' types and characterizations in Section \ref{sec:asset}. 
Subsequently, we examine some standard contingent claims bubbles in Section \ref{sec:contingent}.
Finally, Section \ref{sec:conclude} concludes the main result in this paper.

\section{Model setting}
\label{sec:basic}

\subsection{Financial market}
We consider a discrete sample space with $K\in\mathbb{Z}_{+}$ states $\Omega=\{\omega_1,\cdots, \omega_K\}$, where $\mathbb{Z}_{+}$ denotes the set of positive integers.
To model Knightian uncertainty in the financial market, we introduce a family of probability measures $\mathcal{P}$ defined on $\Omega$ to describe the uncertainty law of the nonlinear randomized trial, see \citet{Yang24subexp} for more details.
Further assume that it satisfies $\sup_{P\in\mathcal{P}}P(\omega)>0$ for all $\omega\in\Omega$.
The market consists of a risky asset and a money market account.
Let $S=(S_t)_{t\ge 0}$ denote the price process of the risky asset, where $S_t\ge 0$ and $S_t$ reflects the ex-dividend price at time $t$. 
Time evolves on the discrete horizon $[0, \bar{T}]$, where $\bar{T}$ is either a finite integer $N$ or $\infty$. Throughout Sections \ref{sec:basic} and \ref{sec:asset}, $t$ and $u$ denote generic observation dates with $t\leq u$, $N$ denotes a deterministic finite bound when such a bound is required, and $\tau$ denotes the liquidation time of the underlying asset. In Section \ref{sec:contingent}, $T$ denotes the fixed maturity of a contingent claim and $\sigma$ denotes an exercise time of an American option. 
Let $\mathcal{F}_t: =\sigma(S_u,\ 0\le u\le t)$ denote the filtration generated by the price process $S=(S_t)_{t\ge 0}$ up to time $t$.
A random time $\tau: \Omega \to [0,\bar{T}] \cup \{\infty\}$ is said to be a $\mathcal{P}$-q.s. stopping time if for all $t\ge 0$, it holds that $\{\tau\le t\} \in\mathcal{F}_t$.
In the following, let $\tau>0$ be a stopping time with respect to the filtration $(\mathcal{F}_t)_{t\ge 0}$, which characterizes the maturity (or default/liquidation time) of the risky assets.
Typical economic events triggering $\tau$ include bankruptcy, acquisition, merger, or antitrust breakup \citep{Protter2013}.
Let $(D_t)_{0\le t<\tau}$ denote the dividend payment process, and let $X_{\tau}$ denote the terminal payoff or liquidation value at time $\tau$, with $D_t\ge 0$ and $X_{\tau}\ge 0$. We use the ex-dividend convention: an investor purchasing the asset at time $t$ receives dividends only at subsequent dates. For notational convenience, at a finite liquidation time we set $D_{\tau}=0$, with the liquidation payment represented by $X_{\tau}$.
The money market account $B=(B_t)_{t\ge 0}$ satisfies $B_0=1$ and 
$$
B_t=(1+r_0)(1+r_1)\cdots(1+r_t),\quad t\ge 1,
$$ 
where $r_u\ge 0$ denotes the spot interest rate over the interval $(u-1, u]$.
Define the discounted wealth process
\begin{equation}
\label{eq:W}
W_t = \hat{S}_t I_{\{t<\tau\}}+ \sum_{u=0}^{t\wedge\tau} \hat{D}_u + \hat{X}_{\tau} I_{\{\tau\le t\}},\quad t\ge 0,
\end{equation}
where $\hat{S}_t = S_t/ B_t,\ \hat{D}_u=D_u/B_u,\ \hat{X}_{\tau}=X_{\tau}/B_{\tau}$. Clearly, $W_t\ge 0$ for all $t$.

A trading strategy consists of an adapted pair of processes $(\pi_t,\eta_t)_{t\ge 0}$, where $\pi_t$ and $\eta_t$ denote the holdings in the risky asset and the money market account from time $t-1$ to $t$, respectively. Short selling of the risky asset is prohibited, so $\pi_k\geq0$, while the money-market position is unrestricted. A strategy is self-financing if its discounted wealth satisfies
\begin{equation}
\label{eq:V-self2}
V_t=V_0+\sum_{k=1}^{t}\pi_k(W_k-W_{k-1}),
\end{equation}
and its discounted portfolio value is
\begin{equation}
\label{eq:V}
V_t=\pi_t\hat S_t I_{\{t<\tau\}}+\eta_t.
\end{equation}
Thus the self-financing condition is expressed through cumulative gains and does not require $V_t=\pi_tW_t$. The nonnegativity of $W$ concerns the underlying asset and its accumulated cash flows. It does not impose nonnegativity on a general portfolio value $V$ or on a contingent claim payoff $H$.

\subsection{No-arbitrage pricing and super-hedging under model uncertainty and short-sale constraints}

Prior to analyzing price bubbles, we first establish the no-arbitrage pricing under model uncertainty and short-sale constraints.
To this end, we introduce the notion of arbitrage under model uncertainty and short-sale constraints.

\begin{definition}
\label{de:arbitrage}
A trading strategy $\pi$ is called an arbitrage over period $[0,u]$ under model uncertainty and short-sale constraints if:

(i) $\pi\ge 0$ is self-financing;

(ii) $V_0=0$;

(iii) $V_{u}(\omega)\ge 0$ for all $\omega\in\Omega$, and $\sup_{P\in\mathcal{P}} E_P[V_{u}]> 0$,\\
where $E_P[\cdot]$ denotes expectation under actual probability measure $P\in\mathcal{P}$.
The market is said to satisfy no-arbitrage under model uncertainty if no trading strategy $\pi$ simultaneously satisfies (i)-(iii).
\end{definition}

\begin{remark}
\label{re:G-ar}
Note that, using the self-financing identity above, (ii) and (iii) in Definition \ref{de:arbitrage} can be reformulated as follows: (ii)' $G_{u}(\omega)\ge 0,\ \forall \omega\in\Omega$ and (iii)' $\sup_{P\in\mathcal{P}} E_P[G_{u}]>0$.
In the absence of model uncertainty, a family of actual probability $\mathcal{P}$ degenerates to a single probability $P$. At this time, Definition \ref{de:arbitrage} becomes the classical no-arbitrage Definition 2.2 in \citet{Pulido14}.
\end{remark}

To ensure absence of arbitrage in markets with short-sale constraints, \citet{Protter2013} introduced a risk neutral probability measure $Q$ satisfying:
\begin{equation}
\label{eq:Q-class}
W_t\ge E_Q[W_{u}\mid\mathcal{F}_t],\quad 0\le t\le u\le\bar T.
\end{equation}
%
%
%
%
%
%
However, model uncertainty further intensifies market incompleteness, which implies that the no-arbitrage asset prices cannot be fully characterized by a single risk neutral probability measure $Q$.
Accordingly, we introduce a family of probability measures $\mathcal{Q}$ for further investigate.
This naturally leads to a family of expectations $\left \{E_Q[\cdot], \ Q\in\mathcal{Q} \right\}$, based on which we construct the strong risk neutral nonlinear expectation in the following form to guarantee no-arbitrage condition.

\begin{definition}
\label{de:Q}
Under model uncertainty and short-sale constraints, $E_Q[\cdot],\ Q\in\mathcal{Q}$ is called a strong risk neutral nonlinear expectation if
\begin{equation}
\label{eq:Q}
W_t\ge \sup_{Q\in\mathcal{Q}} E_Q[W_{u}\mid \mathcal{F}_t],\quad 0\le t\le u\le\bar T,
\end{equation}
where $\sup_{Q\in\mathcal{Q}} Q(\omega)>0$ for all $\omega\in\Omega$.
\end{definition}

\begin{remark}
\label{re:S-G}
Following the terminology of \citet{Peng2019}, the process $W=(W_t)_{t\ge0}$ satisfying (\ref{eq:Q}) is called a $G$-supermartingale.
If the risky asset $\hat{S}$ pays no dividends, then Definition \ref{de:Q} reduces to 
\begin{equation}
\label{eq:S-G}
\hat{S}_t\ge \sup_{Q\in\mathcal{Q}} E_Q[\hat{S}_{u}\mid \mathcal{F}_t],\qquad 0\le t\le u<\tau,
\end{equation}
where $\sup_{Q\in\mathcal{Q}}Q(\omega)>0$ for all $\omega\in\Omega$.
At this time, equation \eqref{eq:S-G} is consistent with Definition 3.4 in \citet{Yang24price}.
When model uncertainty is absent, a family of actual probability $\mathcal{P}$ degenerates to a single probability $P$. Correspondingly, Definition \ref{de:Q} reduces to the classical supermartingale definition, which indicates that \eqref{eq:Q} coincides with \eqref{eq:Q-class}.
\end{remark}

Subsequently, we investigate the relationship between the no-arbitrage condition and the existence of a strong risk neutral nonlinear expectation, and we conclude that the no-arbitrage condition can be derived from the existence of a strong risk neutral nonlinear expectation.

\begin{theorem}
\label{theo:no-ar}
Under model uncertainty and short-sale constraints, the existence of a strong risk neutral nonlinear expectation is sufficient for the no-arbitrage condition.
\end{theorem}

\begin{proof}
Suppose that a self-financing strategy is an arbitrage over $[0,u]$. Then $V_0=0$, $V_u\geq0$, and $V_u(\omega)>0$ for at least one state $\omega$. By the self-financing condition, $V_u=G_u$, where $G_u=\sum_{k=1}^{u}\pi_k(W_k-W_{k-1})$. Since $\pi_k\geq0$ is predictable and \eqref{eq:Q} implies $E_Q[W_k-W_{k-1}\mid\mathcal F_{k-1}]\leq0$ for every $Q\in\mathcal Q$, we have $E_Q[G_u]\leq0$ for every $Q\in\mathcal Q$. On the other hand, $\sup_{Q\in\mathcal Q}Q(\omega)>0$ for every $\omega\in\Omega$ and $V_u$ is nonnegative and strictly positive in at least one state, so $\sup_{Q\in\mathcal Q}E_Q[V_u]>0$. This contradiction proves the no-arbitrage condition.
\end{proof}


We next consider super-hedging over a fixed finite horizon $[0,u]$. Let $H:\Omega\to\mathbb{R}$ be an $\mathcal{F}_u$-measurable terminal payoff, which may take both positive and negative values. Since $\Omega$ is finite, every such real-valued payoff is bounded. A super-hedging strategy consists of the risky asset and the money market account, with a nonnegative stock position and an unrestricted money-market position, and satisfies
\begin{equation}
\label{eq:sup-hedge}
x + \pi\cdot W_0^{u} \ge H,\quad \mathcal{P}-\text{q.s.},
\end{equation}
where $\pi\cdot W_0^{u}:=\sum_{k=1}^{u}\pi_k \Delta W_k = \sum_{k=1}^{u}\pi_k (W_k - W_{k-1})$.
A property holds $\mathcal{P}$-quasi-surely ($\mathcal{P}$-q.s.) if it fails only on a polar set $A$ satisfying $\sup_{P\in\mathcal{P}} P(A) = 0$.
The super-hedging problem aims to determine the initial capital $x$, and subsequently identify the trading strategy $\pi$ by equation (\ref{eq:sup-hedge}).
Prior to characterizing the initial cost, we impose the following absolute continuity condition associated with the strong neutral nonlinear expectation in (\ref{eq:Q}):
\begin{equation}
\label{eq:continuous}
\exists\ P\in\mathcal{P} \text{ such that }Q \ll P,\ \forall\ Q\in\mathcal{Q}.
\end{equation}

\begin{theorem}
\label{theo:sup-re}
Under model uncertainty and short-sale constraints, assume that a strong risk neutral nonlinear expectation exists, the super-hedging price of the contingent claim $H$ is
\begin{align}
\label{eq:sup-re}
\Pi(H):=&\inf\left\{x\in\mathbb{R}:\ \exists\ \pi\ \text{such that}\ x+ \pi\cdot W_0^{u} \ge H,\ \ \mathcal{P}\text{-q.s.} \right\} \\
= &\sup_{Q\in\mathcal{Q}}E_Q[H],\notag \qquad
\end{align}
where $\mathcal{Q}$ is given in Definition \ref{de:Q}. Moreover, $\mathcal{Q}$ satisfies the absolute continuity condition \eqref{eq:continuous}.
\end{theorem}

\begin{proof}
We assume that a strong risk neutral nonlinear expectation exists. Then by Theorem \ref{theo:no-ar} and the sub-additivity of the sublinear expectation yields, 
$$
\sup_{Q\in\mathcal{Q}}E_Q[\pi\cdot W_0^{u}] = \sup_{Q\in\mathcal{Q}}E_Q[ \sum_{k=1}^{u} \pi_k \Delta W_k] \le \sum_{k=1}^{u} \sup_{Q\in\mathcal{Q}}E_Q[\pi_k (W_k-W_{k-1}) ]\le 0,
$$
implying $E_Q[\pi\cdot W_0^{u}]\le 0$ for all $Q\in\mathcal{Q}$.
Taking expectation on both sides of \eqref{eq:sup-hedge}, by the absolute continuity condition \eqref{eq:continuous}, we obtain that $x\ge E_Q[H]$ for all $Q\in\mathcal{Q}$. Hence,
\begin{equation}
\label{eq:theo2}
\Pi(H) = \inf M\ge \sup_{Q\in\mathcal{Q}}E_Q[H],
\end{equation}
where $M$ denotes the set of all initial capitals in \eqref{eq:sup-hedge}. 

Next, we show that all the inequalities in \eqref{eq:theo2} are in fact identities, let $\gamma: = \sup_{Q\in\mathcal{Q}} E_Q[H]$, we need to verify that $\inf M\le \gamma$.
Suppose to the contrary that $\inf M > \gamma$.
Let $C:= \pi \cdot W_0^u$ and $z_{\gamma}:=H-\gamma$. Clearly, $z_{\gamma}\notin C$; otherwise, if $z_{\gamma}\in C$, then $\gamma \in M$, contradicting the assumption $\inf M>\gamma$.
%
By the Hahn-Banach separation theorem yields a separating hyperplane $g$ such that
$$
g\cdot x\le 0,\ \forall x\in C\quad \text{and}\quad g\cdot z_{\gamma}>0.
$$
Define a probability measure $Q$ via the Randon-Nikodym derivative $\frac{dQ}{dP} = g$ for a $P\in\mathcal{P}$.
Since the separating hyperplane is not unique, we can define a set of probability measures and collect them into a set $\mathcal{Q}$.
From $g\cdot x\le 0$ for all $x\in C$, we can deduce $E_Q[C]\le 0$ for all $Q\in\mathcal{Q}$, and thus $\sup_{Q\in \mathcal{Q}} E_Q[W_u]\le W_0$. This implies that $\mathcal{Q}$ satisfies Definition \ref{de:Q} and the absolute continuity condition \eqref{eq:continuous}.
Thus from $g\cdot z_{\gamma}>0$, it follows that
$$
E_Q[H] > \gamma = \sup_{Q\in\mathcal{Q}} E_Q[H],
$$
which contradicts the assumption.
Therefore, we conclude that 
\begin{equation}
	\label{eq:theo2-2}
	\Pi (H) = \inf M\le \gamma = \sup_{Q\in\mathcal{Q}} E_Q[H].
\end{equation}
Combing equations \eqref{eq:theo2} and \eqref{eq:theo2-2} completes the proof.
\end{proof}

The super-hedging cost is invariant under cash translation: for every $a\in\mathbb{R}$,
\[
\Pi(H+a)=\Pi(H)+a.
\]
Indeed, $x+\pi\cdot W_0^u\geq H+a$ if and only if $(x-a)+\pi\cdot W_0^u\geq H$. The stock strategy is unchanged, so the short-sale constraint is preserved. The same argument applies conditionally at each trading node. Consequently, a dual representation established for nonnegative payoffs extends by cash translation to payoffs bounded from below, provided the shifted payoff belongs to its domain.

\begin{remark}
Theorems \ref{theo:no-ar} and \ref{theo:sup-re} establish the no-arbitrage pricing and the super-hedging under model uncertainty and short-sale constraints, which are analogous to the results presented in \citet{Yang24price}.
While in our framework, we consider not only the discounted price process $\hat{S}$, but also the cumulative dividend process $\hat{D}$ and the discounted terminal payoff $\hat{X}$.
\end{remark}

\section{Asset bubbles with short-sale constraints under model uncertainty}
\label{sec:asset}

\subsection{Definition of the asset bubble}

An asset price can be decomposed into two components: its market price, denoted by $\hat{S}=(\hat{S}_t)_{t\ge 0}$, and its fundamental price, denoted by $S^*=(S_t^*)_{t\ge 0}$.
An asset bubble is then identified as the discrepancy between these two quantities \citep{Jarrow2006, Biagini2017}. 
Consequently, a precise specification of the fundamental price is a prerequisite for any rigorous analysis of asset bubbles.
In a classical asset bubble framework, it is commonly to define the fundamental price for a risky asset as the conditional expectation of its discounted future cash flows under an equivalent martingale measure \citep{Jarrow2006}, that is,
\begin{equation}
\label{eq:S*-Q}
S_{t}^*= E_Q\left[\sum_{u=t}^{\tau} \hat{D}_u +  \hat{X}_{\tau}I_{\{\tau< \infty\}}\ \middle| \  \mathcal{F}_t \right] ,\quad 0\le t<\tau.
\end{equation}
To account for the possibility that bubbles may emerge endogenously over time, \citet{Jarrow2010} considered an incomplete financial market framework in which the economy undergoes regime shifts, leading to different local martingale measures across time.
In this setting, the fundamental price accounted for probability measures over partitioned time intervals and aggregated their contributions through summation,
\begin{equation}
\label{eq:S*-Qi}
S_{t}^*= \sum_{i=0}^{\infty} E_{Q^{i}}\left[\sum_{u=t}^{\tau} \hat{D}_u +  \hat{X}_{\tau}I_{\{\tau< \infty\}}\ \middle| \  \mathcal{F}_t \right] I_{\{t\in[\sigma_i,\ \sigma_{i+1})\}} ,\quad 0\le t<\tau,
\end{equation}
where $(\sigma_i)_{i\ge 0}$ denotes an increasing sequence of random times with $\sigma_0=0$ representing the times of regime shifts in the economy.
Motivated by model uncertainty, we instead consider a family of probability measures $\mathcal{Q}$, and evaluate future cash flows via the upper expectation $\sup_{Q\in\mathcal{Q}} E_Q[\cdot]$, which induces a sublinear expectation $\mathbb{E}[\cdot]$ in the sense of \citet{Peng2019}.
Due to its dynamic consistency, the sublinear expectation framework naturally captures the informational structure required for valuation under interval-based specifications of probability models.
Within this setting, we adopt the framework introduced in \cite{Herdegen2016} and \cite{Biagini2017}, defining the fundamental price $S^*$ as the super-hedging price. 
In particular, by Theorem \ref{theo:sup-re}, one can construct a trading strategy that super-replicates the asset's future cash flows.
Under the ex-dividend convention specified in Section \ref{sec:basic}, an investor entering at date $t$ receives dividend payments only at subsequent dates, so the remaining dividend stream below starts at $t+1$.
Accordingly, under short-sale constraints and model uncertainty, the fundamental price of the asset is defined as follows.

\begin{definition}
\label{de:S*}
Under model uncertainty and short-sale constraints, the fundamental price $S^*=(S_t^*)_{t\in [0,\tau)}$ is defined by
\begin{align}
\label{eq:S*-de}
S_t^* : & = \inf \left \{x\in L^0(\mathcal{F}_t):\exists\ \pi \text{ such that } x + \pi \cdot W_t^{\infty} \ge \sum_{u=t+1}^{\tau} \hat{D}_u + \hat{X}_{\tau}I_{\{\tau< \infty\}},\quad  \mathcal{P}-q.s. \right \},\notag\\
& = \sup_{Q\in\mathcal{Q}} E_Q \left[\sum_{u=t+1}^{\tau} \hat{D}_u + \hat{X}_{\tau}I_{\{\tau< \infty\}}\ \middle| \ \mathcal{F}_t \right],\quad 0\le t < \tau,
\end{align}
where $\mathcal{Q}$ is given in Definition \ref{de:Q}. Moreover, $\mathcal{Q}$ satisfies the absolute continuity condition \eqref{eq:continuous}. The infimum is understood conditionally, node by node.
\end{definition}

\begin{remark}
Note that, in Definition \ref{de:S*}, the payoff $\hat{X}_{\tau}$ is not taken into account on the event $\tau=\infty$.
This exclusion is justified by the fact that the payoff $\hat{X}_{\tau} I_{\{\tau = \infty\}}$ cannot be realized through any trading strategy, as all such strategies are required to be liquidated in finite time, even though their time horizons may be unbounded.
\end{remark}

\begin{remark}
The valuation principle in Definition~\ref{de:S*} builds on the superreplication approach to fundamental value; see \citet{Herdegen2016,Biagini2017}. Its implementation here uses the admissible trading strategies and the dual family $\mathcal{Q}$ specified in Section~\ref{sec:basic}. The short-sale restriction enters through these objects. The upper-expectation formula itself is not proposed as a new valuation principle.
\end{remark}


The super-hedging interpretation gives the fundamental price an economic meaning in terms of the cost of covering future cash flows.
Under the super-hedging duality, it is the minimum initial capital required to meet these cash flows using admissible trading strategies.
This interpretation accounts for both the valuation models and the restriction on short positions. The resulting bubble component measures the excess of the market price over this cash-flow benchmark. Such an excess does not by itself provide an admissible arbitrage strategy: selling the asset short and purchasing a cheaper portfolio that covers its cash flows may be prohibited. It is therefore important to distinguish a positive valuation difference from a trading opportunity that investors can actually implement.

We next establish the well-definedness of the fundamental price $S^*$ and investigate the convergence properties of both the fundamental price $S^*$ and the market price $\hat{S}$. 
Before proceeding, we recall several basic concepts from \cite{Denis2011}. 
Let $\mathcal{B}(\Omega)$ denote the Borel $\sigma$-algebra on $\Omega$, and let $L^0(\Omega)$ be the space of all $\mathcal{B}(\Omega)$-measurable real-valued functions. We set,
$$
\mathcal{L}^1 (\mathcal{Q}) := \left \{Y\in L^0(\Omega): \sup_{Q\in\mathcal{Q}} E_Q \left[\left | Y \right | \right]< \infty \right\}.
$$

\begin{lemma}
\label{lemm:S*}
The fundamental price $S^*$ defined in (\ref{eq:S*-de}) is well defined.
Moreover, the market price $(\hat{S}_t)_{t\ge 0}$ converges quasi-surely to a limit $\hat{S}_{\infty}\in \mathcal{L}^1 (\mathcal{Q})$, while the fundamental price $(S^*_t)_{t\ge 0}$ converges quasi-surely to $0$.
\end{lemma}

\begin{proof}
We first show that $S^*$ is well defined.
By definition, it suffices to verify that $\sum_{u=0}^{\tau} \hat{D}_u + \hat{X}_{\tau}I_{\{\tau< \infty\}} \in\mathcal{L}^1 (\mathcal{Q})$ since 
$$
0\le \sum_{u=t+1}^{\tau} \hat{D}_u + \hat{X}_{\tau}I_{\{\tau< \infty\}} \le \sum_{u=0}^{\tau} \hat{D}_u + \hat{X}_{\tau}I_{\{\tau< \infty\}}.
$$
Under the no-arbitrage assumption, the process $(W_t)_{t\ge 0}$ is a nonnegative $G$-supermartingale.
Hence, by the nonlinear supermartingale convergence theorem (see \citet{Peng1999}), there exists $W_{\infty}\in\mathcal{L}^1(\mathcal{Q})$ such that $W_t \to W_{\infty}$, q.s.
Obverse from \eqref{eq:W} that
\begin{equation}
\label{eq:lemm4}
W_{\infty} = \lim_{t\to\infty} W_t = \lim_{t\to\infty} (\hat{S}_t I_{\{t<\tau\}}+ \sum_{u=0}^{t\wedge\tau} \hat{D}_u + \hat{X}_{\tau} I_{\{\tau\le t\}} ) = \lim_{t\to\infty} \hat{S}_t I_{\{\tau = \infty\}} + \sum_{u=0}^{\tau} \hat{D}_u + \hat{X}_{\tau} I_{\{\tau< \infty\}},\quad q.s.
\end{equation}
Since all terms are nonnegative, it follows that
$\hat{S}_{\infty}\in\mathcal{L}^1(\mathcal{Q})$ and $\sum_{u=0}^{\tau} \hat{D}_u + \hat{X}_{\tau} I_{\{\tau< \infty\}} \in\mathcal{L}^1(\mathcal{Q})$, which implies that $S^*$ is well defined and that $\hat{S}_t \to \hat{S}_{\infty}$ quasi-surely.

We now turn to the convergence of $S_t^*$.
By Definition \ref{de:S*}, we observe that
\begin{equation}
\label{eq:lemm1}
\sup_{Q\in\mathcal{Q}}  E_Q \left[ \sum_{u=t+1}^{\tau} \hat{D}_u + \hat{X}_{\tau}I_{\{\tau< \infty\}}\ \middle| \ \mathcal{F}_t \right] =  - \sum_{u=0}^{t} \hat{D}_u + \sup_{Q\in\mathcal{Q}}  E_Q \left[ \sum_{u=0}^{\tau} \hat{D}_u + \hat{X}_{\tau}I_{\{\tau< \infty\}}\ \middle| \ \mathcal{F}_t \right],
\end{equation}
and
\begin{equation}
\label{eq:lemm2}
\sup_{Q\in\mathcal{Q}}  E_Q \left[ \sum_{u=0}^{\tau} \hat{D}_u + \hat{X}_{\tau}I_{\{\tau< \infty\}}\ \middle| \ \mathcal{F}_t \right] = \sup_{Q\in\mathcal{Q}}  E_Q \left[ (\sum_{u=0}^{\tau} \hat{D}_u + \hat{X}_{\tau}) I_{\{\tau< \infty\}} + \sum_{u=0}^{\tau} \hat{D}_u I_{\{\tau= \infty\}} \ \middle| \ \mathcal{F}_t \right].
\end{equation}
Substituting \eqref{eq:lemm2} into \eqref{eq:lemm1} and then into \eqref{eq:S*-de}, and applying the Dominated convergence theorem for sublinear expectation in discrete states \citep{Yang24subexp}, we obtain
\begin{align*}
\lim_{t\to\infty}S_t^* &= -\sum_{u=0}^{\infty} \hat{D}_u I_{\{\tau= \infty\}} + \sup_{Q\in\mathcal{Q}} E_Q \lim_{t\to\infty} \left[ (\sum_{u=0}^{\tau} \hat{D}_u + \hat{X}_{\tau}) I_{\{\tau< \infty\}} + \sum_{u=0}^{\tau} \hat{D}_u I_{\{\tau= \infty\}} \ \middle| \ \mathcal{F}_t \right]I_{\{\tau= \infty\}}\\
& = -\sum_{u=0}^{\infty} \hat{D}_u I_{\{\tau= \infty\}} + \sup_{Q\in\mathcal{Q}} E_Q \left[\sum_{u=0}^{\infty} \hat{D}_u  \ \middle| \ \mathcal{F}_{\infty} \right]I_{\{\tau= \infty\}} = 0,\quad q.s.
\end{align*}
This completes the proof.
\end{proof}

Given the definition of the fundamental price process $S^*$, the associated fundamental wealth process $W^*=(W_t^*)_{t\ge 0}$ is naturally induced by the wealth process in \eqref{eq:W}. More precisely, it is given by
\begin{equation}
\label{eq:W*}
W_{t}^* = S^*_t I_{\{t<\tau\}}+ \sum_{u=0}^{t\wedge\tau} \hat{D}_u + \hat{X}_{\tau}I_{\{\tau\le t\}}.
\end{equation}
We then investigate the convergence properties of $W^*$ and establish its G-martingale structure.

\begin{lemma}
\label{lemm:W*}
The fundamental wealth process $(W_t^*)_{t\ge 0}$ is a uniformly integrable G-martingale.
Moreover, it is closed by a terminal value $W_{\infty}^*\in\mathcal{L}^1(\mathcal{Q})$.
\end{lemma}

\begin{proof}
We first identify the terminal value of $W^*$.
By Lemma \ref{lemm:S*}, we have $S_t^* \to 0$ quasi-surely as $t\to\infty$.
Consequently,
\begin{equation}
\label{eq:lemm3}
W_{\infty}^* = \lim_{t\to\infty} W_{t}^* =  \lim_{t\to\infty} (S^*_t I_{\{t<\tau\}}+ \sum_{u=0}^{t\wedge\tau} \hat{D}_u + \hat{X}_{\tau}I_{\{\tau\le t\}})=\sum_{u=0}^{\tau} \hat{D}_u + \hat{X}_{\tau}I_{\{\tau< \infty\}},\quad q.s.
\end{equation}
From Lemma \ref{lemm:S*}, it follows that $W^*_{\infty}\in\mathcal{L}^1(\mathcal{Q})$.
We now verify the $G$-martingale property.
Recall equations \eqref{eq:lemm4} and \eqref{eq:lemm3}, it is obviously that $W_{\infty}= W_{\infty}^* + S_{\infty}$.
Thus, we obtain that
\begin{align*}
\sup_{Q\in\mathcal{Q}} E_Q[W_{\infty}^* \mid \mathcal{F}_t] &= \sup_{Q\in\mathcal{Q}} E_Q \left[\sum_{u=0}^{\tau} \hat{D}_u I_{\{\tau\le t\}} + \hat{X}_{\tau} I_{\{\tau\le t\}} + \sum_{u=0}^{t} \hat{D}_u I_{\{t< \tau\}} + \sum_{u=t+1}^{\tau} \hat{D}_u I_{\{t< \tau\}} + \hat{X}_{\tau} I_{\{t<\tau< \infty\}} \ \middle| \ \mathcal{F}_t \right] \\
& = \sup_{Q\in\mathcal{Q}} E_Q \left[\sum_{u=t+1}^{\tau} \hat{D}_u + \hat{X}_{\tau} I_{\{\tau<\infty\}} \ \middle| \ \mathcal{F}_t \right] I_{\{t<\tau\}} + \left ( \sum_{u=0}^{t}\hat{D}_u I_{\{t<\tau\}} + \sum_{u=0}^{\tau}\hat{D}_u I_{\{\tau\le t\}} \right) + \hat{X}_{\tau} I_{\{\tau\le t\}} \\
& = S_{t}^*I_{\{t<\tau\}} + \sum_{u=0}^{t\wedge\tau}\hat{D}_u + \hat{X}_{\tau}I_{\{\tau\le t\}} = W_t^*.
\end{align*}
Hence, $(W_t^*)_{t\ge 0}$ is a $G$-martingale.
Finally, since $W_t^* = \sup_{Q\in\mathcal{Q}} E_Q[W_{\infty}^* \mid \mathcal{F}_t]$ and $W_{\infty} \in \mathcal{L}^1(\mathcal{Q})$, it follows that $(W_t^*)_{t\ge 0}$ is uniformly integrable and closed by $W_{\infty}^*$.
This completes the proof.
\end{proof}

After introducing the notion of the fundamental price and establishing its main properties, we now formalize the concept of an asset price bubble. 
\begin{definition}
\label{de:bubb}
The asset price bubble $\beta=(\beta_t)_{t\ge 0}$ is defined by
\begin{equation}
\label{eq:beta}
\beta_t := \hat{S}_t -S^*_t,
\end{equation}
where $\hat{S}_t$ denotes the market price of asset, and $S_t^*$ denotes its fundamental price as defined in (\ref{eq:S*-de}).
\end{definition}

\begin{remark}
\label{re:W-W*}
The bubble definition in \eqref{eq:beta} coincides with the standard notion in the literature; see, for instance, \citet{Jarrow2006, Jarrow2010, Protter2013, Herdegen2016}. 
Combining the representations of the wealth process $W$ in \eqref{eq:W} and the fundamental wealth process $W^*$  in \eqref{eq:W*}, the bubble process can be equivalently written as
$$
\beta_t = W_t - W_t^*, \quad t<\tau.
$$
Thus, the asset price bubble can be interpreted as the difference between the market wealth process and its fundamental wealth.
The absence of bubbles corresponds to $\beta_t=0$ for all $t\ge 0$, $\mathcal{P}$-q.s. A bubble exists if there is some $t\ge 0$ such that
\begin{equation}
	\label{eq:beta>0}
	\sup_{Q\in \mathcal{Q}} Q(\beta_t >0)>0,
\end{equation}
where $\sup_{Q\in \mathcal{Q}} Q(\omega)>0$ for all $\omega\in\Omega$.
Equation \ref{eq:beta>0} means that there exists a probability measure $Q\in\mathcal{Q}$ under which $\beta_t>0$ holds with strictly positive probability. This notion of bubble existence is consistent with that in \citet{Biagini2017}.
\end{remark}

\begin{remark}
	\label{re:bubble-strong_E}
	It is important to note that, under model uncertainty and short-sale constraints, the super-hedging duality theorem in Theorem \ref{theo:sup-re} relies on the existence of a strong risk neutral nonlinear expectation. The definition of the fundamental price $S^*$ in Definition \ref{de:S*} is based on the same assumption. This assumption is natural and well justified in the context of bubble analysis considered in this paper.
	Indeed, \citet{Protter2013} argued that, in equity markets, asset price bubbles typically satisfy $\beta_t\ge 0$. In other markets, such as foreign exchange markets, negative bubbles may arise; however, a systematic analysis of such cases is beyond the scope of the present paper and is left for future research.
	Motivated by this observation, within our framework of model uncertainty and short-sale constraints, this leads to
	\begin{equation}
		\label{eq:S>S*}
		\hat{S}_t \ge S_t^* = \sup_{Q\in\mathcal{Q}} E_Q \left[ \sum_{u=t+1}^{\tau} \hat{D}_u + \hat{X}_{\tau} I_{\{\tau<\infty\}} \ \middle| \ \mathcal{F}_t \right], \quad 0\le t<\tau,
	\end{equation}
	Combing \eqref{eq:S>S*} with the wealth process \eqref{eq:W} yields
	$$
	W_t \ge \sup_{Q\in\mathcal{Q}} E_Q[W_{\infty} \mid \mathcal{F}_t],\quad t<\tau< \infty.
	$$
	Therefore, the assumption of a strong risk-neutral nonlinear expectation is consistent with the economic structure of the market and can be viewed as a natural and economically justified modeling assumption under model uncertainty and short-sale constraints.
\end{remark}

\subsection{Bubble dynamics under uncertainty}

Following the definition of the asset price bubble, we examine the classification types of bubbles under the condition of model uncertainty and short-sale constraints.
In the financial market without model uncertainty and short-sale constraints, \citet{Jarrow2006, Jarrow2010} classified bubbles in into three types according to the behavior of a stopping time $\tau$: the case $P(\tau =\infty)>0$, the case where $\tau$ is unbounded but satisfies $P(\tau<\infty)=1$, and the case where $\tau$ is bounded.
Under model uncertainty, upper and lower conditional expectations provide different bounds on the evolution of a bubble. We distinguish these bounds when studying the effect of the liquidation-time assumptions. For brevity, we use the term infi-supermartingale for the lower conditional expectation inequality defined below.

\begin{definition}
Let $Y=(Y_t)_{t\ge 0}$ denotes a random variable sequence on a discrete state space $\Omega$. Then $Y$ is called a infi-supermartingale if it satisfies
\begin{equation}
\label{eq:inf-supmar}
Y_t\ge \inf_{Q\in\mathcal{Q}} E_Q[Y_u \mid\mathcal{F}_t],\quad 0\le t\le u,
\end{equation}
where $\sup_{Q\in\mathcal{Q}}Q(\omega)>0$ for all $\omega\in \Omega$.
\end{definition}

\begin{remark}
\label{re:sup-inf}
Under the standing integrability and conditional-expectation conventions,
\[
\inf_{Q\in\mathcal{Q}}E_Q[Y_u\mid\mathcal{F}_t]
\leq
\sup_{Q\in\mathcal{Q}}E_Q[Y_u\mid\mathcal{F}_t].
\]
Hence every $G$-supermartingale is also an infi-supermartingale. These are nested conditions, rather than mutually exclusive classes of processes.

The $G$-supermartingale condition bounds the future conditional expectation under every admissible valuation measure. The infi-supermartingale condition bounds only the lower envelope of these expectations; it does not require the supermartingale inequality to hold under every measure, or under one fixed measure for all dates.

If the valuation family $\mathcal{Q}$ consists of a single measure, both conditions reduce to the usual supermartingale condition under that measure.
\end{remark}

The distinction matters for the persistence of a bubble after it has disappeared. For a nonnegative bubble satisfying the upper conditional expectation bound, a zero current value forces its future value to remain zero under every admissible valuation measure. The lower bound alone gives no such guarantee: a low valuation under some models does not exclude positive future valuations under others. The two inequalities therefore differ in the conclusions they support about bubble disappearance and possible subsequent emergence.

\begin{theorem}
\label{theo:bubb-pro}
Under the stated market assumptions, the bubble process satisfies the following conditional expectation bounds in the respective liquidation-time cases:

(i) $\beta_t$ is a infi-supermartingale, if the stopping time $\tau$ is unbounded with $P(\tau<\infty) = 1$, or satisfies $P(\tau=\infty)>0$.

(ii) $\beta_t$ is a $G$-supermartingale, if the stopping time $\tau$ is bounded.
\end{theorem}

\begin{proof}
We begin with the proof of type (ii). Assume that $\tau\leq N$ for some deterministic finite $N$, then $S_{\infty}=0$. Combing equations \eqref{eq:lemm4} and \eqref{eq:lemm3}, it is obviously that $W_{\infty} = W_{\infty}^* + S_{\infty}$. Thus $W_{\infty} = W_{\infty}^*$.
Define the auxiliary process 
\begin{equation}
\label{eq:theo8}
\bar{\beta}_t := W_t - \sup_{Q\in\mathcal{Q}} E_Q[W_{\infty}\mid \mathcal{F}_t].
\end{equation}
Since $W_t^* = \sup_{Q\in\mathcal{Q}} E_Q[W_{\infty}^* \mid \mathcal{F}_t]$ by Lemma \ref{lemm:W*}, it follows that
$$
\label{eq:theo6}
\beta_t  = W_t - W_t^* = \bar{\beta}_t + \sup_{Q\in\mathcal{Q}} E_Q[W_{\infty} \mid \mathcal{F}_t] -W_t^* = \bar{\beta}_t + \sup_{Q\in\mathcal{Q}} E_Q[W_{\infty}^* \mid \mathcal{F}_t] -W_t^* = \bar{\beta}_t.
$$
Choose $N\geq\tau$; then $\beta_N=\bar{\beta}_N=0$.
Suppose, to the contrary, that $\beta_t$ is not a $G$-supermartingale. Then, it follow that
$$
\beta_t < \sup_{Q\in \mathcal{Q}} E_Q[\beta_u \mid \mathcal{F}_t] = 0,
$$
which contradicts the standing assumption that $\beta_t \ge 0$ (see Remark \ref{re:bubble-strong_E}).
Consequently, $\beta_t$ is a $G$-supermartingale. This established assertion (ii).

When $\tau$ is unbounded but satisfies $P(\tau<\infty)=1$, we still have $S_{\infty}=0$. Consequently, $\beta_t = \bar{\beta}_t$ holds.
By combining Lemma \ref{lemm:W*} with equations \eqref{eq:Q} and \eqref{eq:theo8}, we obtain
\begin{align*}
\sup_{Q\in\mathcal{Q}} E_Q[-\bar{\beta}_u \mid \mathcal{F}_t] & = \sup_{Q\in\mathcal{Q}} E_Q[\sup_{Q\in\mathcal{Q}}E_Q[W_{\infty} \mid \mathcal{F}_u] - W_u \mid \mathcal{F}_t] \ge \sup_{Q\in\mathcal{Q}} E_Q[W_{\infty} \mid \mathcal{F}_t] - \sup_{Q\in\mathcal{Q}} E_Q[W_u \mid \mathcal{F}_t] \\
& \ge \sup_{Q\in\mathcal{Q}} E_Q[W_{\infty} \mid \mathcal{F}_t] -W_t = -\bar{\beta}_t,\qquad 0\le t\le u.
\end{align*}
Hence $\bar{\beta}_t \ge \inf_{Q\in\mathcal{Q}} E_Q[\bar{\beta}_u \mid \mathcal{F}_t]$, that is, $\beta_t$ satisfies infi-supermartingale.

If $P(\tau =\infty)>0$, then by Remark \ref{re:W-W*}, Lemma \ref{lemm:W*} and equation \eqref{eq:Q}, we obtain similarly that
\begin{align*}
\sup_{Q\in\mathcal{Q}} E_Q[-\beta_u \mid \mathcal{F}_t]&= \sup_{Q\in\mathcal{Q}} E_Q[W_u^* - W_u \mid \mathcal{F}_t] \ge \sup_{Q\in\mathcal{Q}} E_Q[W_u^* \mid \mathcal{F}_t] - \sup_{Q\in\mathcal{Q}} E_Q[W_u \mid \mathcal{F}_t]\\
& \ge W_t^* - W_t =-\beta_t,\qquad 0\le t\le u,
\end{align*}
indicating $\beta_t$ is also an infi-supermartingale.
This completes the proof.
\end{proof}

\begin{remark}
The conditional expectation bounds in Theorem \ref{theo:bubb-pro} differ from the classical three-case framework of \citet{Jarrow2010}.
Specifically, when $P(\tau=\infty)>0$, \citet{Jarrow2010} showed that the bubble process $\beta_t$ is a uniformly integrable local martingale, whereas if $\tau$ unbounded with $P(\tau<\infty)=1$, then $\beta$ is merely a local martingale.
However, under model uncertainty and short-sale constraints, both cases can be naturally captured by the notion of an infi-supermartingale. Indeed, for any $Q\in\mathcal{Q}$, we have
$$
\beta_t = E_Q[\beta_u\mid \mathcal{F}_t]\ge \inf_{Q\in\mathcal{Q}} E_Q[\beta_u \mid \mathcal{F}_t],\quad 0\le t\le u.
$$
Thus, the infi-supermartingale inequality provides a common lower conditional expectation bound for these two unbounded-lifetime cases under model uncertainty.
If $\tau$ is a bounded stopping time, \citet{Jarrow2010} established that $\beta_t$ is a strict local martingale, and hence a supermartingale. 
Under the presence of model uncertainty and short-sale constraints, however, $\beta_t$ must be characterized by the stronger condition of being a $G$-supermartingale.
If the valuation family $\mathcal{Q}$ consists of a single measure, the concept of a $G$-supermartingale coincides with the classical notion of a supermartingale under that measure.
\end{remark}

As indicated in Theorem \ref{theo:bubb-pro}, Type 1 bubbles arise when an asset has a finite but unbounded lifespan, or an infinite lifespan with a payoff at $\{\tau=\infty\}$. 
Type 2 bubbles pertain to assets with a bounded lifespan. 
Subsequently, we present corresponding examples for different types of stopping time $\tau$. 
Following \cite{Jarrow2006}, we take fiat money, declared by government as legal tender, as one such example.

\begin{example}
\label{exam:tau-infty}
Let $\hat{S}$ denote fiat money. In the benchmark case of a perfectly stable unit of account, we normalize $\hat{S}_t=1$ for all $t\ge 0$. 
Under financial markets with model uncertainty, however, it is inappropriate to model the real value of fiat money as constant.
We therefore assume that the discounted real value of fiat money is given by 
$$
\hat{S}_t = \frac{1}{P_t},\quad t\ge 0,
$$
where $P_t$ denotes a price index (e.g., the consumer price index)
We normalize the initial price level to $P_0 =1$ and assume that the price index evolves according to $P_{t+1} = P_t Z_{t+1}$, where $Z_{t+1}$ is the one-period inflation factor.
Model uncertainty is introduced by allowing $Z_{t+1} \in [\underline{z}, \overline{z}],\ \underline{z}< 1\le \overline{z}$, thereby capturing the possibility of both deflation and inflation arising from macroeconomic uncertainty.
As before, we assume that fiat money pays no dividends, i.e., $D=0$, which implies that the fundamental value satisfies $S_t^* =0$ for all $t\ge 0$.
Consequently, the entire value of fiat money is attributed to the bubble component $\beta_t = \hat{S}_t$.
For any $t\ge 0$, we obtain
$$
\inf_{Q\in\mathcal{Q}} E_Q[\beta_{t+1} \mid \mathcal{F}_t] = \inf_{Q\in\mathcal{Q}} E_Q \left[\frac{1}{ P_t Z_{t+1}} \ \middle | \ \mathcal{F}_t \right] = \beta_t \frac{1}{\overline{z}} \le \beta_t.
$$
We conclude that the bubble process $(\beta_t)_{t\ge 0}$ is an infi-supermartingale with respect to the family of probability measures $\mathcal{Q}$.
\end{example}

To address the case $\tau<\infty$, including situations in which $\tau$ is unbounded but satisfies $P(\tau<\infty)=1$, as well as the case of bounded maturity, we consider a simpler and explicit example in a discrete-time setting.
We begin with the former scenario.

\begin{example}
\label{exam:tau-unbounded}
Let $\Omega=\{\omega_1, \omega_2\}$, and let the asset maturity $\tau$ be a strictly positive random time such that $P(\tau >t)>0$ for all $t\ge 0$.
Assume the asset pays no dividends and delivers a unit payoff at maturity $\tau$. 
Then, by (\ref{eq:S*-de}), the fundamental price process is given by $(S_t^*)_{t\ge 0} = (I_{\{t<\tau\}})_{t\ge 0}$.
We define a discounted price process $\hat{S}$ on a one-step tree and compute the corresponding bubble process $\beta$:
$$
\hat{S}_0 =2 
\begin{matrix}
& \nearrow & \hat{S}_1(\omega_1)=2.5  & \to & \beta_1(\omega_1) =  1.5\\
& \searrow & \hat{S}_1(\omega_2)=1.5  & \to & \beta_1(\omega_2) =  0.5
\end{matrix}
$$
Following \citet{Yang24subexp}, we characterize the probability set $\mathcal{Q}$ by a convex, closed domain $\mathcal{D}=\{\theta_1: 0.2\le \theta_1 \le 0.4\}$, where $\theta_1 = Q(\omega_1)$.
It then follows that $\beta_0=1$ and
$$
\inf_{Q\in\mathcal{Q}} E_Q[\beta_1] = \inf_{\theta_1 \in [0.2, 0.4]} [0.5\ \theta_1 -0.5\ (1-\theta_1)] = 0.7.
$$
Consequently, $\beta_0> \inf_{Q\in\mathcal{Q}} E_Q[\beta_1]$.
By contrast, if $\mathcal{D}=\{\theta_1: 0.5\le \theta_1 \le 0.7\}$, then $\beta_0 = \inf_{Q\in\mathcal{Q}} E_Q[\beta_1] =1$.
In both cases, we obtain $\beta_t \ge \inf_{Q\in\mathcal{Q}} E_Q[\beta_u \mid \mathcal{F}_t]$ for $t\le u$, illustrating the infi-supermartingale property of the bubble process $(\beta_t)_{t\ge 0}$.
\end{example}

Analogously to Example \ref{exam:tau-unbounded}, we now consider the case in which the maturity $\tau$ is bounded.

\begin{example}
In the same discrete framework as in Example \ref{exam:tau-unbounded}, we define the fundamental price $S^*$ and the discounted market price $\hat{S}$ on the space $\Omega=\{\omega_1, \omega_2\}$.
Assume that $\tau\leq N$ for some deterministic finite $N$, and let the probability set $\mathcal{Q}$ be represented by a convex closed region $\mathcal{D}=\{\theta_1: 0.2\le \theta_1 \le 0.4\}$ where $\theta_1 = Q(\omega_1)$. 
As before, we have $\beta_0=1$. A direct computation yields
$$
\sup_{Q\in\mathcal{Q}} E_Q[\beta_1] = \sup_{\theta_1 \in [0.2, 0.4]} [0.5\ \theta_1 -0.5\ (1-\theta_1)] = 0.9.
$$
Consequently, $\beta_0> \sup_{Q\in\mathcal{Q}} E_Q[\beta_1]$.
If instead $\mathcal{D}=\{\theta_1: 0.2\le \theta_1 \le 0.5\}$, then $\beta_0 = \sup_{Q\in\mathcal{Q}} E_Q[\beta_1] =1$.
Therefore, in the case of bounded maturity $\tau$, we obtain $\beta_t \ge \sup_{Q\in\mathcal{Q}} E_Q[\beta_u \mid \mathcal{F}_t]$ for $t\le u$, which illustrates the $G$-supermartingale property of the bubble process $(\beta_t)_{t\ge 0}$.
\end{example}

\subsection{Necessary and sufficient conditions for the existence of bubbles}

Subsequently, We investigate necessary and sufficient conditions for the existence of  price bubbles under model uncertainty and short-sale constraints.
Our objective is to characterize, based solely on the observed asset price process $\hat{S}$, whether a bubble is present in the market.
Theorem~\ref{theo:bubb-pro} relates the liquidation-time assumptions to conditional expectation bounds on the bubble process. The definition of the bubble remains the same across these cases.
We therefore begin by providing a necessary condition for the existence of a nontrivial bubble under short-sale constraints and model uncertainty.

\begin{theorem}
\label{theo:bubb-S-tau}
Under model uncertainty and short-sale constraints, if there exists a nontrivial bubble $\beta$ in the asset price, then we have two possibilities of the discounted market price $\hat{S}$ under model uncertainty:

(i) $\hat{S}_t$ is a infi-supermartingale if the stopping time $\tau$ is unbounded with $P(\tau<\infty)=1$, or satisfies $P(\tau=\infty)>0$.

(ii) $\hat{S}_t$ is a $G$-supermartingale if the stopping time $\tau$ is bounded.
\end{theorem}

\begin{proof}
We begin with the proof of (i). Since $\hat{D}_t\ge 0$ for all $t\ge 0$, it follows directly from equation \eqref{eq:S*-de} that
\begin{equation}
\label{eq:theo3}
\sup_{Q\in\mathcal{Q}} E_Q[S_u^* \mid \mathcal{F}_t] = \sup_{Q\in\mathcal{Q}} E_Q \left[ \sum_{k=t+1}^{\tau} \hat{D}_k -\sum_{k=t+1}^{u} \hat{D}_k + \hat{X}_{\tau} I_{\{\tau< \infty\}}\ \middle| \ \mathcal{F}_t \right]\le S_t^*,\ \  0\le t\le u.
\end{equation}
Combining \eqref{eq:theo3} with Theorem \ref{theo:bubb-pro}, we obtain 
\begin{align*}
\hat{S}_t = \beta_t + S_t^* &\ge \inf_{Q\in\mathcal{Q}} E_Q[\beta_u \mid \mathcal{F}_t] + \sup_{Q\in\mathcal{Q}} E_Q[S_u^* \mid \mathcal{F}_t]=  \inf_{Q\in\mathcal{Q}} E_Q[\beta_u \mid \mathcal{F}_t] - \inf_{Q\in\mathcal{Q}} E_Q[-S_u^* \mid \mathcal{F}_t]\\
& \ge \inf_{Q\in\mathcal{Q}} E_Q[\beta_u+S_u^* \mid \mathcal{F}_t] =\inf_{Q\in\mathcal{Q}} E_Q[\hat{S}_u \mid \mathcal{F}_t],\qquad 0\le t\le u,
\end{align*}
where $\sup_{Q\in\mathcal{Q}} Q(\omega)>0$ for all $\omega\in \Omega$.
This shows that the discounted market price process $(\hat{S}_t)_{t\ge 0}$ is an infi-supermartingale whenever $\tau$ is unbounded with $P(\tau<\infty)=1$, or when $P(\tau=\infty)>0$.
We now turn to (ii), if the stopping time $\tau$ is bounded, then Theorem \ref{theo:bubb-pro} together with (\ref{eq:theo3}) yields
$$
\hat{S}_t = \beta_t + S_t^* \ge \sup_{Q\in\mathcal{Q}} E_Q[\beta_u \mid \mathcal{F}_t] + \sup_{Q\in\mathcal{Q}} E_Q[S_u^* \mid \mathcal{F}_t]\ge \sup_{Q\in\mathcal{Q}} E_Q[\hat{S}_u \mid \mathcal{F}_t],\quad 0\le t\le u,
$$
where $\sup_{Q\in\mathcal{Q}} Q(\omega)>0$ for all $\omega\in\Omega$.
Hence, $(\hat{S}_t)_{t\ge 0}$ is a $G$-supermartingale when $\tau$ is bounded. 
This completes the proof.
\end{proof}

We next examine which martingale properties of the market price process allow for the emergence of a bubble.
In particular, we establish sufficient conditions for the existence of a bubble under short-sale constraints and model uncertainty.
For convenience, we impose the simplifying assumption that the asset $\hat{S}$ pays no dividends.

\begin{theorem}
\label{theo:bubb-S-G}
Under model uncertainty and short-sale constraints, assume that the discounted asset price process $\hat{S}= (\hat{S}_t)_{t\ge 0}$ pays no dividends.
For any stopping time $\tau$, if $\hat{S}$ is a $G$-supermartingale, then a bubble exists. 
\end{theorem}

\begin{proof}
Under the no dividends assumption, assume that the discounted asset price process $\hat{S}$ is a $G$-supermartingale.
Suppose, by contradiction, that no price bubble exists; that is, $\beta_t = 0$ for all $t\ge 0$.
At the terminal time $\tau$, the asset is liquidated and delivers the payoff $X_\tau$, so that its ex-dividend price satisfies $\hat{S}_{\tau} = X_{\tau}$ on $\{\tau < \infty\}$.
Then we have
$$
\hat{S}_t = S_t^* = \sup_{Q\in\mathcal{Q}} E_Q \left[\hat{X}_{\tau}I_{\{\tau< \infty\}}\ \middle | \ \mathcal{F}_t \right] = \sup_{Q\in\mathcal{Q}} E_Q[\hat{S}_{\tau}\mid \mathcal{F}_t],\quad 0\le t\le \tau.
$$
This shows that $\hat{S}_t$ is a $G$-martingale, which contradicts the assumption that $\hat{S}_t$ is a $G$-supermartingale. Therefore, a price bubble must exist.
This completes the proof.
\end{proof}

\begin{remark}
In the absence of model uncertainty, the notion of a $G$-supermartingale in equation (\ref{eq:S-G}) degenerates into a classical supermartingale.
Accordingly, Theorem \ref{theo:bubb-S-G} recovers the classical result of \citet{Protter2013}.
\end{remark}

\begin{remark}
\label{re:NA-no bubble}
By Remark \ref{re:bubble-strong_E}, it is natural and economically justified in our framework to assume the existence of a strong risk neutral nonlinear expectation.
Then by Theorem \ref{theo:no-ar}, the financial market is arbitrage-free.
Moreover, in view of Theorem \ref{theo:bubb-S-G}, any non-dividend paying asset whose price is a $G$-supermartingale necessarily admits a bubble.
Consequently, under model uncertainty and short-sale constraints, if the discounted asset price $\hat{S}$ of non-dividend-paying asset is a $G$-supermartingale, then the market is arbitrage-free and the asset price exhibits a bubble.
\end{remark}

Combining Theorems \ref{theo:bubb-S-tau} and \ref{theo:bubb-S-G}, we conclude that when the stopping time $\tau$ is bounded and the asset $\hat{S}$ pays no dividends, the $G$-supermartingle property of $\hat{S}$ provides both necessary and sufficient conditions for the existence of a bubble under model uncertainty and short-sale constraints.
In contrast, when $\tau$ is unbounded with $P(\tau<\infty)=1$ or satisfies $P(\tau=\infty)>0$, the infi-supermartingale property of $\hat{S}$ is no longer sufficient to ensure the presence of a bubble; in this case, the condition must be strengthened to the $G$-supermartingale framework.
For an intuitive illustration of these results, Figure \ref{fig:Bubble} summarizes the martingale properties of the price process and the corresponding necessary and sufficient conditions for the existence of bubbles under model uncertainty and short-sale constraints.
Subsequently, we summarize the properties of bubbles in the following corollary.

\begin{corollary}
\label{coro:bubb-T}
Under model uncertainty and short-sale constraints, assume that the asset $\hat{S}$ pays no dividends.
Then any nontrivial asset price bubble $\beta$ satisfies the following properties:

(i) $\beta\ge 0$.

(ii) $\beta_{\tau} I_{\{\tau<\infty\}}= 0$.

(iii) If the stopping time $\tau$ is bounded and $\beta_t =0$, then $\beta_u=0$ for all $u\ge t$. 
In contrast, this implication generally fails for other types of stopping times, such as unbounded $\tau$ with $P(\tau <\infty)=1$ or $\tau$ satisfying $P(\tau=\infty)>0$.
\end{corollary}

\begin{proof}
For (i), under the non-dividend assumption, Remark \ref{re:S-G} shows that the existence of a strong risk neutral nonlinear expectation implies that $\hat{S}$ is a $G$-supermartingale. Consequently, for any $t<\tau$, $\hat{S}_t \ge \sup_{Q\in \mathcal{Q}} E_Q[\hat{S}_{\tau} \mid \mathcal{F}_t]$. Since $\hat{S}_{\tau} = X_{\tau}$ on $\{\tau <\infty\}$, it follows that
$$
\beta_t = \hat{S}_t - \sup_{Q\in \mathcal{Q}} E_Q[X_{\tau} I_{\{\tau <\infty\}} \mid \mathcal{F}_t] = \hat{S}_t - \sup_{Q\in\mathcal{Q}} E_Q[\hat{S}_{\tau} \mid \mathcal{F}_t] \ge 0,
$$
which verifies the claim.
For (ii), on the event $\{\tau <\infty\}$, equation \eqref{eq:S*-de} yields
$$
\beta_{\tau} = \hat{S}_{\tau} - S_{\tau}^* = \hat{S}_{\tau} - \sup_{Q\in\mathcal{Q}} E_Q[\hat{X}_{\tau} I_{\{\tau< \infty\}} \mid \mathcal{F}_{\tau}] = 0.
$$
which proves the claim.
For (iii), suppose first that $\tau$ is bounded and that $\beta_t = 0$. By Theorem \ref{theo:bubb-pro}, we have $0\ge \sup_{Q\in\mathcal{Q}}E_Q[\beta_u \mid \mathcal{F}_t]$ for $u\ge t$. By the nonnegativity of $\beta_u\ge 0$ in (i), it follows that
$$
\sup_{Q\in\mathcal{Q}}E_Q[\beta_u \mid \mathcal{F}_t]= 0,\quad 0\le t\le u.
$$
Because $\sup_{Q\in\mathcal{Q}}Q(\omega)>0$ for every $\omega\in\Omega$, we conclude that $\beta_u =0$ for all $u\ge t$.
If, on the other hand, $\tau$ is not bounded, then Theorem \ref{theo:bubb-pro} together with $\beta_t=0$ only implies
$$
\inf_{Q\in\mathcal{Q}} E_Q[\beta_u \mid \mathcal{F}_t] =0,\quad 0\le t\le u.
$$
However, since $\sup_{Q\in\mathcal{Q}}Q(\omega)>0$ for all $\omega\in \Omega$, this condition is insufficient to deduce that $\beta_u =0$. Hence,the persistence property established in the bounded case does not extend to unbounded stopping times.
This completes the proof.
\end{proof}

Property (i) of Corollary~\ref{coro:bubb-T} shows that the bubble component is nonnegative, so that the market price of the asset is no lower than its fundamental value under the maintained assumptions. Property (ii) further implies that the bubble vanishes at a finite maturity or liquidation time, since the asset's value is then settled by its terminal payoff. Moreover, property (iii) shows that, when the stopping time $\tau$ is bounded, a zero-bubble state is persistent: if $\beta_t=0$ at some time $t$, then $\beta_u=0$ for all $u\geq t$. Thus, in the bounded-maturity case, a bubble cannot emerge after a zero-bubble state has been reached; in particular, if no bubble is present initially, no bubble can arise subsequently, and once an existing bubble disappears, it cannot reappear. This persistence property follows from the $G$-supermartingale structure of the bubble process together with its nonnegativity, rather than from a particular specification of investor behavior. Economically, bounded maturity imposes a terminal date by which the asset's value must ultimately be resolved through its cash flows, placing a stronger restriction on admissible bubble dynamics.
In contrast, when $\tau$ is unbounded, either with $P(\tau<\infty)=1$ or with $P(\tau=\infty)>0$, the weaker infi-supermartingale condition does not imply the same persistence property. Therefore, in these cases, the present result does not rule out the possibility that a bubble may re-emerge after reaching zero.

\subsection{No dominance under model uncertainty}

In the preceding sections, Remarks \ref{re:bubble-strong_E} and \ref{re:NA-no bubble} indicate that, in the absence of dividends, the financial market is arbitrage-free and a price bubble may arise under our framework. We next show that no dominance rules out bubbles when the liquidation time is bounded.

\begin{definition}
\label{de:dominant}
Fix two finite dates $0\leq t<u$. An admissible self-financing strategy dominates another admissible self-financing strategy over $[t,u]$ if their discounted wealth processes $V$ and $\widetilde V$ satisfy
\[
V_t=\widetilde V_t,\qquad V_u\geq \widetilde V_u \quad \mathcal P\text{-q.s.},
\]
and
\[
\sup_{P\in\mathcal P}P(V_u>\widetilde V_u)>0.
\]
Both strategies must satisfy the short-sale constraint on the underlying asset. The market satisfies no dominance if no such pair exists on any finite comparison interval $[t,u]$.
\end{definition}

\begin{remark}
\label{re:dominant-G}
For portfolios with the same value at time $t$, the terminal wealth comparison is equivalent to the comparison of their cumulative discounted gains. If a single prior is specified, the quasi-sure comparison reduces to an almost-sure comparison under that prior; this does not imply that the family of dual pricing measures is a singleton.
\end{remark}

\begin{remark}
An arbitrage strategy dominates the zero-investment strategy. Consequently, no dominance implies no arbitrage. The converse need not hold under short-sale constraints because the difference between two admissible portfolios need not itself be an admissible trading strategy.
\end{remark}

\begin{theorem}
\label{theo:no dominant}
Suppose that $\tau\leq N$ $\mathcal P$-q.s. for some deterministic finite $N$. Under model uncertainty and short-sale constraints, if the no-dominance condition holds, then
\[
\hat S_t=S_t^*,\qquad \beta_t=0,\qquad 0\leq t<\tau,\quad \mathcal P\text{-q.s.}
\]
\end{theorem}

\begin{proof}
Fix $t<N$ and consider $\{t<\tau\}$. Since $\tau\leq N$, all remaining cash flows are received by $N$. By the super-hedging definition in Definition \ref{de:S*}, $S_t^*$ is the infimum of the initial capitals needed to cover their accumulated discounted value at $N$. Buying and holding one unit of the asset, with all subsequent distributions retained in the money market account, delivers this value. Hence $S_t^*\leq\hat S_t$.

Suppose that $S_t^*<\hat S_t$ on a nonpolar $\mathcal F_t$-measurable event. Since the state space is finite, choose an $\mathcal F_t$-atom contained in this event. By the definition of the infimum, there is an admissible super-hedging strategy on this atom with initial capital $x<\hat S_t$.

Investing the difference $\hat S_t-x$ in the money market account gives the same initial wealth as the buy-and-hold strategy and strictly greater wealth at $N$. Outside the atom, use the buy-and-hold strategy. This contradicts no dominance on $[t,N]$. Therefore $\hat S_t=S_t^*$ quasi-surely on $\{t<\tau\}$, and $\beta_t=0$.
\end{proof}

\begin{remark}
The boundedness of $\tau$ ensures that all remaining cash flows are realized by a common finite date $N$. The super-hedging definition of the fundamental value can therefore be compared directly with buy-and-hold wealth on $[t,N]$. This argument does not require market completeness, but it does not extend directly to an unbounded liquidation time under the present definition of no dominance on finite trading intervals.
\end{remark}

\section{Contingent claims bubbles under model uncertainty and short-sale constraints}
\label{sec:contingent}

Section \ref{sec:asset} examines the difference between the market price of the underlying asset and its fundamental value. We now turn to contingent claims whose payoffs depend on that market price. Forward contracts and European call and put options are useful because their terminal payoffs satisfy an exact linear identity, while American options introduce the additional feature of early exercise. These contracts allow us to compare fundamental values with market prices and to examine how short-sale constraints and model uncertainty affect the corresponding valuation relations.

For the contracts considered below, let $T$ be a fixed finite maturity date. We assume that the underlying asset pays no dividends over $[0,T]$ and that its liquidation time satisfies $\tau>T$ $\mathcal P$-q.s., so the contracts mature while the underlying asset remains in existence. For this finite-horizon analysis, the terminal value of holding the underlying is its discounted resale value $\hat S_T$. We retain the notation $S_t^*$ and $\beta_t$, suppressing their dependence on the fixed horizon $T$, and define
\begin{equation}
\label{eq:S*-T}
S_t^*=\sup_{Q\in\mathcal Q}E_Q[\hat S_T\mid\mathcal F_t],\qquad \beta_t=\hat S_t-S_t^*,\qquad 0\leq t\leq T.
\end{equation}
This is the finite-horizon valuation used in this section; it is not obtained from the lifetime cash-flow definition solely from the condition $\tau>T$.

Throughout this section, we assume that the conditional super-hedging duality corresponding to Theorem \ref{theo:sup-re} holds on the fixed horizon $[0,T]$, so that the fundamental values defined by conditional upper expectations coincide with the corresponding minimum admissible super-hedging costs. The hedging instruments used for these fundamental valuations are the underlying asset and the money market account; the contingent claims being valued are not included as additional hedging instruments. The stock position remains nonnegative, whereas the money-market position is unrestricted. Market prices of the contingent claims are considered in the enlarged market, where the claims may be traded in both long and short positions; whenever no dominance is invoked, it refers to this enlarged market.

For a European-type claim with discounted terminal payoff $H_T(\hat S)$, let $\Lambda_t(H)$ denote its discounted market price. Its fundamental value is
\[
\Lambda_t^*(H)=\sup_{Q\in\mathcal Q}E_Q[H_T(\hat S)\mid\mathcal F_t],\qquad 0\leq t\leq T,
\]
and we write $\delta_t=\Lambda_t(H)-\Lambda_t^*(H)$ for the difference between its market and fundamental prices. Unlike the nonnegative underlying component $\beta_t$, $\delta_t$ is not assumed to be nonnegative.

\subsection{Forward contracts and European call and put options}

Subsequently, we examine some standard contingent claims written on the same risky asset: a forward contract, a European call option, and a European put option.
All quantities are expressed in discounted terms. The delivery price $K\geq0$ is a fixed constant expressed in units of the money market account.
A forward contract with delivery price $K$ and maturity $T$ yields the discounted terminal payoff $\hat{S}_T - K$, its discounted market price at time $t$ is denoted by $F_t(K)$.
A European call option with strike $K$ and maturity $T$ pays the discounted payoff $(\hat{S}_T - K)^{+}$, with discounted market price $C_t^E(K)$ at time $t$.
Similarly, a European put option with strike $K$ and maturity $T$ delivers the discounted payoff $(K-\hat{S}_T)^{+}$, and its discounted market price at time $t$ is denoted by $P_t^E(K)$.
Let $F_t^*(K)$, $C_t^{E*}(K)$, and $P_t^{E*}(K)$ denote the corresponding fundamental prices of the forward contract, the European call option, and the European put option, respectively. These are defined by
\begin{align*}
F_t^*(K) &= \sup_{Q\in\mathcal{Q}} E_Q[\hat{S}_T - K \mid \mathcal{F}_t],\\
C_t^{E*}(K) &= \sup_{Q\in\mathcal{Q}} E_Q[(\hat{S}_T - K)^+ \mid \mathcal{F}_t],\\ 
P_t^{E*}(K) &= \sup_{Q\in\mathcal{Q}} E_Q[(K - \hat{S}_T)^+ \mid \mathcal{F}_t],
\end{align*}
where $\sup_{Q\in\mathcal{Q}} Q(\omega)>0$ for all $\omega\in \Omega$. The forward payoff may be negative, but it is bounded from below by $-K$ because $\hat{S}_T\geq0$. Adding $K$ units of discounted cash converts this payoff into the nonnegative payoff $\hat{S}_T$. Cash translation therefore yields
\[
F_t^*(K)=\sup_{Q\in\mathcal Q}E_Q[\hat S_T\mid\mathcal F_t]-K=S_t^*-K.
\]
The corresponding super-hedging portfolios differ only in their money-market positions, so this translation preserves the short-sale constraint on the underlying asset.

We next compare the upper valuations of these three terminal payoffs. Although the payoffs satisfy the usual put-call identity, valuing each payoff separately by its minimum constrained super-hedging cost need not preserve this identity. The following result gives the corresponding bounds. These are relations between valuation benchmarks rather than a statement that observed option prices violate put-call parity.

\begin{theorem}
\label{theo:C*-P*}
Under model uncertainty and short-sale constraints, the fundamental prices of European options and forward contract satisfy
\begin{align}
\label{eq:C*-P*}
\inf_{Q\in\mathcal{Q}} E_Q[\hat{S}_T - K \mid \mathcal{F}_t] \le C_t^{E*}(K) - P_t^{E*}(K) & \le \sup_{Q\in\mathcal{Q}} E_Q[\hat{S}_T - K \mid \mathcal{F}_t]\\
& = S_t^* - K = F_t^*(K) \notag ,\quad 0\le t\le T,
\end{align}
where $\sup_{Q\in\mathcal{Q}} Q(\omega)>0$ for all $\omega\in \Omega$.
\end{theorem}

\begin{proof}
At maturity $T$, the payoffs satisfy the identity
$$
(\hat{S}_T - K)^{+} - (K-\hat{S}_T)^{+} = \hat{S}_T - K.
$$
Using the sub-additivity of the sublinear expectation, we obtain
\begin{align*}
C_t^{E*}(K) - P_t^{E*}(K) & = \sup_{Q\in\mathcal{Q}} E_Q[(\hat{S}_T - K)^{+}\mid \mathcal{F}_t] - \sup_{Q\in\mathcal{Q}} E_Q[(K - \hat{S}_T)^{+}\mid \mathcal{F}_t]\\
& \le \sup_{Q\in\mathcal{Q}} E_Q[\hat{S}_T - K \mid \mathcal{F}_t] = S_t^* - K = F_t^*(K).
\end{align*}
Similarly, we have
\begin{align*}
C_t^{E*}(K) - P_t^{E*}(K) & = -\inf_{Q\in\mathcal{Q}} E_Q[-(\hat{S}_T - K)^{+}\mid \mathcal{F}_t] + \inf_{Q\in\mathcal{Q}} E_Q[- (K - \hat{S}_T)^{+}\mid \mathcal{F}_t]  \\
& \ge \inf_{Q\in\mathcal{Q}} E_Q[\hat{S}_T - K \mid \mathcal{F}_t].
\end{align*}
This completes the proof.
\end{proof}

\begin{remark}
Theorem \ref{theo:C*-P*} follows from the subadditivity of the conditional upper expectation. Its role is to describe the relation between separate upper valuations; it does not provide a criterion for the existence of an asset price bubble.

For each fixed $Q\in\mathcal Q$, linearity gives
\[
E_Q[(\hat S_T-K)^+\mid\mathcal F_t]-E_Q[(K-\hat S_T)^+\mid\mathcal F_t]=E_Q[\hat S_T-K\mid\mathcal F_t].
\]
Taking the upper valuation of each payoff separately need not preserve this equality. Under the super-hedging interpretation, the resulting quantities are the costs of covering different liabilities using the permitted trading strategies.

The inequalities in Theorem \ref{theo:C*-P*} need not be strict. In particular, suppose that, for some $\mathcal F_t$-measurable random variable $m_t$,
\[
E_Q[\hat S_T\mid\mathcal F_t]=m_t\qquad\text{for every }Q\in\mathcal Q.
\]
Then
\[
C_t^{E*}(K)-P_t^{E*}(K)=m_t-K=F_t^*(K).
\]
Indeed, under every $Q$, the conditional call expectation equals the conditional put expectation plus the same quantity $m_t-K$, so taking suprema preserves the equality. Thus, multiple valuation measures do not by themselves imply a put-call disparity.

A singleton valuation family is a special case in which parity holds. However, the absence of model uncertainty does not by itself make the super-hedging dual family a singleton, since trading constraints or market incompleteness may still generate multiple dual measures.
\end{remark}


\begin{example}
Consider a one-period market with zero interest, an initial underlying price $\hat S_0=100$, and terminal prices $\hat S_1\in\{80,120\}$. Trading in the underlying is subject to a short-sale prohibition, while borrowing and lending through the money market account are unrestricted. 
The contingent claims below are valued by super-hedging with these two assets; they are not added to the set of hedging instruments.

Let $q$ denote the probability of the upper state under a dual measure $Q_q$. The supermartingale restriction is
\[
E^{Q_q}[\hat S_1]
=80(1-q)+120q\leq100,
\]
so the full dual set is
\[
\mathcal Q=\{Q_q:0\leq q\leq\tfrac12\}.
\]
For strike $K=100$, the upper valuations of the European call,
European put, and forward payoff are
\[
\begin{aligned}
	C_0^{E*}(100)
	&=\sup_{0\leq q\leq1/2}20q=10,\\
	P_0^{E*}(100)
	&=\sup_{0\leq q\leq1/2}20(1-q)=20,\\
	F_0^*(100)
	&=\sup_{0\leq q\leq1/2}(40q-20)=0.
\end{aligned}
\]
Consequently,
\[
C_0^{E*}(100)-P_0^{E*}(100)
=-10<0=F_0^*(100).
\]

These values also have a direct hedging interpretation. The call is replicated by holding one half of a share and borrowing $40$, at an initial cost of $10$. The put is super-hedged by holding $20$ in cash. Indeed, if $x$ is the initial capital and $h\geq0$ is the stock holding, covering the put payoff in the lower state requires
\[
x+h(80-100)\geq20,
\]
and hence $x\geq20+20h\geq20$. The cash strategy is therefore optimal. The forward payoff is replicated by holding one share and borrowing $100$, at zero initial cost.
If short selling of the underlying is permitted, the put can instead be replicated by holding minus one half of a share and $60$ in cash, at an initial cost of $10$. The call and forward values remain unchanged. Thus,
\[
\begin{array}{lccc}
	\hline
	\text{Trading regime}
	& C_0^{E*}(100) & P_0^{E*}(100) & F_0^*(100)\\
	\hline
	\text{Short selling permitted} & 10 & 10 & 0\\
	\text{Short selling prohibited} & 10 & 20 & 0\\
	\hline
\end{array}
\]

The strict inequality in this example has a direct hedging explanation. Replicating the put requires a short stock position, which is excluded by the constraint, so the least costly admissible alternative is more expensive. By contrast, the replicating strategies for the call and the forward already satisfy the constraint, so their costs remain unchanged. The difference between the call and put upper valuations changes because the restriction affects their hedging costs differently.

This disparity occurs without a bubble in the underlying under the finite-horizon definition used in this section. Indeed,
\[
S_0^*=\sup_{0\leq q\leq1/2}\bigl[80(1-q)+120q\bigr]=100=\hat S_0,
\]
so that $\beta_0=0$, whereas
\[
C_0^{E*}(100)-P_0^{E*}(100)=-10<0=F_0^*(100).
\]
Thus, the disparity between these upper valuations is not, by itself, evidence of an underlying asset price bubble.

The values in the table are separate super-hedging costs using the stock and the money market account. They are not simultaneous market quotes for freely tradable claims. In particular, the upper cost of covering a put liability is not automatically an amount that can be received by selling the put. The example therefore illustrates the valuation effect of a trading restriction rather than an executable arbitrage based on observed option prices.

\end{example}

We now turn from the separate super-hedging valuations to the market prices of the traded contracts. The valuation disparity illustrated above does not determine whether these market prices satisfy put-call parity. The following result addresses this separate question under no dominance in the enlarged market.

\begin{theorem}
Assume that no dominance holds at every trading date in the enlarged market containing the underlying asset, the money market account, and the European options and forward contract under consideration.
Then the put-call parity holds for market prices, that is,
\begin{equation}
\label{eq:C-P=F}
C_t^E(K)-P_t^E(K)=\hat S_t-K=F_t(K).
\end{equation}
\end{theorem}

\begin{proof}
The following three portfolios have the same discounted terminal payoff $\hat S_T-K$: a long European call combined with a short European put, a long forward contract, and one unit of the underlying asset combined with a money-market position of $-K$. All three portfolios are admissible. If two of them had different prices, purchasing the cheaper portfolio and investing the price difference in the money market account would dominate the more expensive portfolio with the same initial wealth. No dominance therefore implies
\[
C_t^E(K)-P_t^E(K)=F_t(K)=\hat S_t-K.
\]
\end{proof}

Combining the upper-valuation bounds with market put-call parity yields restrictions on the differences between market prices and fundamental values.

\begin{theorem}
\label{theo:call}
Under the finite-horizon assumptions of this section and no dominance in the enlarged market,
\[
\beta_t=\delta_t^F=0,\qquad
\delta_t^{EC}-\delta_t^{EP}\geq0,\qquad 0\leq t\leq T.
\]
\end{theorem}

\begin{proof}
Holding one unit of the underlying asset until $T$ delivers $\hat S_T$. By the conditional super-hedging duality stated at the beginning of this section, $S_t^*$ is the minimum cost of super-hedging this payoff. If $S_t^*<\hat S_t$ on a nonpolar event, a cheaper super-hedge together with the remaining cash would dominate buy-and-hold over $[t,T]$. Thus, by the same argument as in the proof of Theorem \ref{theo:no dominant},
\[
\hat S_t=S_t^*,\qquad \beta_t=0.
\]
Using the market-price relation above and cash translation invariance gives
\[
F_t(K)=\hat S_t-K=S_t^*-K=F_t^*(K),
\]
so that $\delta_t^F=0$.
Finally, Theorem \ref{theo:C*-P*} and equation \eqref{eq:C-P=F} yield
\[
\begin{aligned}
\delta_t^{EC}-\delta_t^{EP}
&=\bigl(C_t^E(K)-P_t^E(K)\bigr)-\bigl(C_t^{E*}(K)-P_t^{E*}(K)\bigr)\\
&=F_t^*(K)-\bigl(C_t^{E*}(K)-P_t^{E*}(K)\bigr)\geq0.
\end{aligned}
\]
This completes the proof.
\end{proof}

\begin{remark}
Theorem \ref{theo:call} is a no-dominance benchmark. Under its assumptions, the underlying and forward valuation differences vanish. The inequality for the call-put valuation differences reflects the nonlinear aggregation of expectations in the fundamental prices and does not establish the existence of a positive option-price bubble.

Moreover, when a European claim can be sold short and its fundamental value is its minimum admissible super-hedging cost, no arbitrage implies
\[
\Lambda_t(H)\leq\Lambda_t^*(H).
\]
Otherwise, selling the claim and purchasing a cheaper super-hedge would generate an arbitrage. Thus the quantities $\delta_t$ are differences between market and fundamental prices and should not automatically be interpreted as positive speculative bubbles.
\end{remark}

\subsection{American options}

After analyzing forward contracts and European options, we proceed to study American options under short-sale constraints and model uncertainty. Let $\mathcal T_{t,T}$ denote the set of stopping times taking values in $\{t,t+1,\ldots,T\}$. We use $\sigma$ for an exercise time, reserving $\tau$ for the liquidation time of the underlying asset. Taking into account the time value of money, we define the fundamental price of an American option with payoff function $H$ and maturity $T$ as
\[
\Lambda_t^{A*}(H)=\sup_{\sigma\in\mathcal T_{t,T}}\sup_{Q\in\mathcal Q}E_Q[H_\sigma(\hat S)\mid\mathcal F_t].
\]
Let $C_t^{A*}(K)$ and $P_t^{A*}(K)$ denote the fundamental prices of American call and put options with strike $K$, respectively. Then
\[
C_t^{A*}(K)=\sup_{\sigma\in\mathcal T_{t,T}}\sup_{Q\in\mathcal Q}E_Q[(\hat S_\sigma-K)^+\mid\mathcal F_t],
\]
\[
P_t^{A*}(K)=\sup_{\sigma\in\mathcal T_{t,T}}\sup_{Q\in\mathcal Q}E_Q[(K-\hat S_\sigma)^+\mid\mathcal F_t].
\]
The corresponding differences between market and fundamental prices for the American call and put options are defined by
\[
\delta_t^{AC}=C_t^A(K)-C_t^{A*}(K),\qquad
\delta_t^{AP}=P_t^A(K)-P_t^{A*}(K),
\]
where $C_t^A(K)$ and $P_t^A(K)$ denote the respective market prices. We first examine the relationship between American and European call options.

\begin{theorem}
\label{theo:C^A-E}
Under the finite-horizon assumptions of this section and no dominance in the enlarged market, American and European call options with the same strike and maturity satisfy
\[
C_t^{A*}(K)=C_t^{E*}(K),\qquad 0\leq t\leq T.
\]
If both options are dynamically traded and may be sold short, their market prices also satisfy
\[
C_t^A(K)=C_t^E(K).
\]
Consequently,
\[
\delta_t^{AC}=\delta_t^{EC}.
\]
\end{theorem}

\begin{proof}
By Theorem \ref{theo:call} and equation \eqref{eq:S*-T},
\[
\hat S_u=S_u^*=\sup_{Q\in\mathcal Q}E_Q[\hat S_T\mid\mathcal F_u],\qquad 0\leq u\leq T.
\]
Since time is discrete and $T$ is finite, this identity also holds at every $\sigma\in\mathcal T_{t,T}$. Monotonicity and cash translation invariance imply
\[
(\hat S_\sigma-K)^+\leq\sup_{Q\in\mathcal Q}E_Q[(\hat S_T-K)^+\mid\mathcal F_\sigma].
\]
Taking the conditional upper expectation at $t$ and using time consistency gives
\[
\sup_{Q\in\mathcal Q}E_Q[(\hat S_\sigma-K)^+\mid\mathcal F_t]\leq C_t^{E*}(K).
\]
Taking the supremum over $\sigma$ yields $C_t^{A*}(K)\leq C_t^{E*}(K)$, while the reverse inequality follows by choosing $\sigma=T$.

For market prices, no arbitrage gives nonnegative European call and put prices. Together with equation \eqref{eq:C-P=F}, this implies
\[
C_u^E(K)\geq(\hat S_u-K)^+,\qquad 0\leq u\leq T.
\]
If $C_t^A(K)>C_t^E(K)$, sell the American call and buy the European call. At any exercise time of the American call, the European call can be sold for at least the exercise payment, so the strictly positive initial price difference gives an arbitrage. Conversely, if $C_t^A(K)<C_t^E(K)$, buy the American call, sell the European call, and exercise the American call at $T$; the terminal payoffs cancel, again giving an arbitrage. Hence $C_t^A(K)=C_t^E(K)$, and combining the two price equalities gives $\delta_t^{AC}=\delta_t^{EC}$.
\end{proof}

Theorem \ref{theo:C^A-E} identifies the no-dominance benchmark for call options in the present dividend-free setting. Once the underlying asset price equals the minimum cost of super-hedging its terminal resale value, early exercise does not increase the fundamental value of a call option; under the stated trading assumptions, the same conclusion holds for market prices. Accordingly, the result should not be interpreted as a positive early-exercise premium generated by an underlying bubble, since the underlying component is zero under these assumptions. Model uncertainty may still affect individual valuations and the relation between upper call and put values, but it does not by itself establish a nonzero bubble-related call premium in this benchmark.

\subsection{Economic and financial implications}

The distinction between fundamental values and market prices is important for interpreting overvaluation in a constrained market.
An investor may be able to cover an asset's future cash flows at a cost below its market price, yet be unable to profit from the difference because the required short position is prohibited.
The relevant question is therefore not only whether a price difference exists, but also whether an admissible strategy can exploit it.
This distinction explains why a cash-flow valuation benchmark and a no-arbitrage condition serve different purposes in the analysis.

For valuation and hedging, the framework emphasizes the role of the admissible trading strategies. A portfolio that replicates a payoff when short positions are allowed may cease to be available when the underlying cannot be shorted. 
As the one-period example illustrates, this can raise the cost of covering a put payoff while leaving the corresponding call cost unchanged. The effect depends on the payoff and the positions required to hedge it; a trading restriction need not affect all contingent claims in the same way.

The option results also require a distinction between upper valuations and observed market quotes. 
The terminal payoffs of a call, a put, and a forward satisfy a linear identity, but taking an upper expectation separately for each payoff need not preserve that identity. Different probability measures can determine the upper valuations of different claims. A gap between these valuations should therefore not be interpreted automatically as evidence of a market pricing inconsistency or a bubble. Such an interpretation requires additional information about actual quotes, trading opportunities, and the valuation family.

For market monitoring, the analysis suggests that an assessment of price deviations should state both the cash-flow valuation benchmark and the trading restrictions under which the comparison is made. Changes in the model specification or in the ability to take short positions can alter the interpretation of a valuation difference. 
The present framework does not determine whether introducing or removing a short-sale restriction improves welfare or market stability, since investor preferences, market clearing, and policy responses are not modeled. Its contribution is to clarify the valuation and trading conditions that must be distinguished before drawing such conclusions.

\section{Conclusion}
\label{sec:conclude}

In this paper, we investigate discrete-time asset price bubbles under model uncertainty and short-sale constraints.
For a wealth process consisting of the market price of a risky asset together with accumulated cash flows, we establish a fundamental theorem of asset pricing under model uncertainty and short-sale constraints. In particular, we show that the no-arbitrage condition is guaranteed by the $G$-supermartingale property of the wealth process.
The super-hedging representation provides the valuation benchmark used to define the asset's fundamental value and its bubble component in the constrained market.
We verify that this fundamental price is well defined and analyze its convergence properties. 
We examine upper and lower conditional expectation bounds for the bubble process under different liquidation-time assumptions. The upper bound is stronger and, together with nonnegativity, implies that a bubble cannot reappear after reaching zero. The lower bound alone does not imply this persistence property.
We further derive necessary and sufficient conditions for the existence of bubbles.
When maturity is bounded and the asset pays no dividends, the $G$-supermartingale property of the asset price is both necessary and sufficient for the existence of bubbles.
In contrast, when maturity is unbounded, the infi-supermartingale property yields a necessary condition, while the $G$-supermartingale property becomes sufficient.
When the liquidation time is bounded, no dominance rules out bubbles because the lifetime cash-flow fundamental value can be compared directly with the terminal wealth of a buy-and-hold strategy. For finite-horizon contingent claims, we instead use the underlying asset's resale value at the option maturity as the terminal valuation object. Within this setting, fundamental values satisfy put-call bounds, while market prices satisfy put-call parity under no dominance. The underlying and forward valuation differences then vanish. American and European call fundamental values coincide, and their market prices also coincide under the stated trading assumptions. These conclusions identify the restrictions imposed by no dominance; they do not establish a positive bubble-driven early-exercise premium.

The financial interpretation of this paper rests on separating cash-flow valuation, admissible trading strategies, and observed market prices. In particular, an upper valuation difference is not automatically an executable trading profit. The numerical illustration shows how a short-sale restriction can change hedging costs and the relationship between separately evaluated contingent claims.
An empirical application would require a specification of the valuation family, data on asset and option prices, and an account of the relevant trading restrictions. It would also need to distinguish valuation differences generated by model uncertainty and trading constraints from those associated with transaction costs or other market frictions. Developing and assessing such an empirical implementation is left for future research.

\bibliography{gexpffb}

\end{document}